\documentclass[hidelinks,11pt,reprint,showpacs,showkeys,nofootinbib,ssfamily]{article}
\usepackage{axodraw2}
\usepackage{amsmath,fontenc,calrsfs}
\usepackage{amsfonts,amssymb,mathtools,slashed,empheq,psfrag,epsfig}
\usepackage{cite} 
\usepackage[breaklinks]{hyperref} 
\hypersetup{backref, colorlinks=blue  } 
\usepackage{multirow}
\usepackage{float}
\usepackage{appendix}
\usepackage{color}
\usepackage{url}
\usepackage{subfigure}
\usepackage{footnote}
\usepackage{authblk} 

\newcommand{\mk}{|\vk|}

\newcommand{\val}{\vec{\alpha}}
\newcommand{\vall}{\mathbf{a}}

\newcommand{\cT}{N} 

\newcommand{\vp}{{\mathbf{p}}}
\newcommand{\vq}{{\mathbf{q}}}

\newcommand{\vQ}{{\mathbf{Q}}}

\newcommand{\vk}{{\mathbf{k}}}

\newcommand{\Opd}{{\mathcal{O}}(p^2)}

\newcommand{\hatr}{\hat{\rho}}

\newcommand{\bg}{\begin{align}}
\newcommand{\eeg}{\end{align}}
\newcommand{\be}{\begin{equation}}
\newcommand{\ee}{\end{equation}}
\newcommand{\ba}{\begin{eqnarray}}
\newcommand{\ea}{\end{eqnarray}}

\newcommand{\nn}{\nonumber}

\newcommand{\barr}[1]{\not\mathrel #1}

\newcommand{\ep}{\epsilon}

\newcommand{\vs}{\vspace{-0.0cm}}

\newcommand{\vz}{\mathbf{z}}

\begin{document}

\thispagestyle{empty}

\vspace{2cm}

\begin{center}
{\Large{\bf An in-medium chiral power counting for nuclear matter and some applications}}
\end{center}
\vspace{.5cm}

\begin{center}
{\Large J.~A. Oller}
\vskip 10pt
{\it Departamento de F\'{\i}sica, Universidad de Murcia, E-30071 Murcia, Spain}\\
\end{center}

\vspace{1cm}
\noindent
\begin{abstract}
We review on a chiral power counting scheme for in-medium chiral perturbation theory with 
nucleons and 
 pions as explicit  degrees of freedom coupled to external sources. 
It allows for a systematic expansion taking into account both local as well as pion-mediated inter-nucleon interactions. 
One can identify from this power counting classes of non-perturbative diagrams that require resummation. 
A non-perturbative method based on 
Unitary Chiral Perturbation Theory (UChPT) was also developed for performing those needed resummations. 
 This power counting and non-perturbative techniques were firstly applied to calculate 
the pion self-energy, the pion-decay constants and the quark condensate
 in  nuclear matter up-to-and-including next-to-leading order (NLO) contributions. 
 The cancellation of the contributions  at NLO to the pion self-energy and decay constants from  
in-medium nucleon-nucleon ($NN$) interactions was derived. 
Some NLO contributions from the in-medium $NN$ interactions survive for the quark condensate 
due to the quark-mass dependence of the pion mass. 
Next, we discuss the calculation of the energy density in the nuclear medium by employing the 
derived in-medium $NN$ scattering amplitudes. 
For symmetric and  neutron matter it reproduces  in good agreement, and without fine tuning,  calculations  from 
realistic $NN$ potentials with a model for the three-nucleon interaction. 
These results are applied to derive the equation of state (EOS) 
for neutron stars and obtain an upper limit for a neutron mass slightly above 2 solar masses, in agreement with 
recent observations. Furthermore, our results also fulfill other constraints 
from the detection of the gravitational waves in the event GW170817 by the LIGO and Virgo Collaborations, like the upper bound
on  the maximal mass of a neutron star and the allowed interval of values for the radius of a 
$1.4$-solar-mass neutron star. 
The knowledge of the neutron-matter EOS is also employed to give an upper bound of the gravitational 
constant within the strong gravitational field of a 2 solar-mass neutron star.
\end{abstract} 
\noindent{\it Keywords\/}: chiral power counting, non-perturbative calculations, nuclear matter, neutron stars, gravitational waves

\newpage

\tableofcontents

\newpage

\section{Introduction}
\label{sec:int}

Nuclear physics treats typically systems of many nucleons. 
One of the long standing issues in nuclear physics is the calculation of atomic nuclei and nuclear-matter properties 
from microscopic internucleon forces in a systematic and controlled way.
In recent decades Effective Field Theory (EFT) has proven to be an important tool to accomplish that goal. 
It is based on a power counting that establishes a hierarchy between the infinite amount of contributions, 
so that a finite number of mechanisms have to be considered for an order given. 
In this way, a controlled expansion results that allows  to estimate the expected error due to the truncation of the series. 
In this work we employ  Chiral Perturbation Theory (ChPT) \cite{wein,wein1,wein2} to nuclear systems,
with  the nucleons and the pions being the degrees of freedom. 
ChPT  has also the virtue of being connected with QCD, since it shares the same symmetries and breaking of them. 

In Ref.\cite{prcoller} many-body field theory was derived from quantum field theory by considering nuclear matter as a continuous 
set of free nucleons filling a Fermi sea at asymptotic times. 
The generating functional of ChPT in the presence of external sources was deduced, 
similarly as in the pion and pion-nucleon sectors \cite{gl1,sainio}. 
These results were applied in Ref.\cite{annp} to study ChPT in nuclear matter but including only perturbative nucleon interactions 
due to pion exchanges.
Thus, the contact (short-range) nucleon-nucleon (and multi-nucleon) interactions were neglected. 
Other works along these lines are Refs.~\cite{jido.181228.1,jido.181228.2}, which work out the next-to-leading order (NLO) 
in-medium corrections to
 some pion properties by applying the many-body formalism derived in Ref.~\cite{prcoller} and
following the perturbative power counting of Ref.~\cite{annp}.
As in the latter reference, the detailed analysis of a pion state in the nuclear medium within this formalism allows Ref.~\cite{jido.181228.1} to single out the essential role played by the pion wave-function renormalization in defining the in-medium pion coupling
constants. Furthermore, as first derived in Ref.~\cite{annp}, the in-medium pion wave-function renormalization gives account
of the so-called missing $S$-wave  repulsion \cite{oset.181228.1,kaiser.181228.1}.

Let us stress that, as it is well known since the seminal papers of Weinberg \cite{wein1,wein2}, 
the nucleon propagators do not always count as $1/p$, being $p$ a typical low momentum much smaller 
than the chiral scale of the expansion.
It is often the case that the nucleon propagators scale as the inverse of a nucleon kinetic energy, $m/p^2$ (with $m$
the physical nucleon mass), 
so that they are much larger than assumed. 
This, of course, invalidates the straightforward application of the pion-nucleon 
power counting in vacuum.
These issues were overcome in Ref.~\cite{lacour.180827.2} where an in-medium power counting that takes into account contact,
as well pion-mediated nucleon interactions,
together with the possible infrared enhancements from the in-medium nucleon propagators was developed. 

The in-medium power counting of Ref.~\cite{lacour.180827.2}
has been applied in the literature to several problems of interest, like the 
in-medium pion self-energy and the energy per nucleon  of nuclear matter \cite{lacour.180827.3}, 
the in-medium quark condensate and the pion-decay constants \cite{lacour.180827.4}. 
The resulting equation of state (EOS) for pure neutron matter of Ref.~\cite{lacour.180827.3} was used in Ref.~\cite{llanes.180827.1}
to calculate the upper bound of a neutron star mass, which turned out to be above two solar masses,
as required by the recent detection in Ref.~\cite{demorest.180827.1} of a two-solar mass neutron star.
We also notice here that it follows from the EOS of Ref.~\cite{lacour.180827.3}
that the radius of a neutron star of $1.4$~solar masses should be around 12.6~km \cite{llanes.180827.1},
well inside the interval of values allowed by the analysis of 
 Ref.~\cite{annala.180827.1} based on the GW170817 event detected by the LIGO and Virgo 
 Collaborations \cite{ligo.180827.1} and on the 
existence of the two-solar mass neutron star detected in Ref.~\cite{demorest.180827.1}.
We review on these results here.

After this introduction, we rederive in Sec.~\ref{sec:pw} the chiral power counting in the nuclear 
medium of Ref.~\cite{lacour.180827.2} that takes into account
 multi-nucleon local interactions, pion exchanges and the possible infrared enhancement of nucleon propagators.
We then discuss the vacuum and in-medium nucleon-nucleon ($NN$) interactions used for the calculations  up to 
NLO of in-medium ChPT, Sec.~\ref{sec:nn-int}. 
This theory is then applied to calculate the EOS of nuclear matter in 
Secs.~\ref{sec.180829.1} and \ref{sec.180829.1b}, 
which is used to the study of neutron-star properties, and then connecting with the GW170817 observations as well.
 An important point to characterize the state  of nuclear matter  is to study sufficient and/or necessary conditions for the 
 breaking of chiral symmetry. In this respect a discussion on the results of 
Refs.~\cite{annp,lacour.180827.2,lacour.180827.3,lacour.180827.4}
on the temporal pion decay constant $f_t$ (corresponding to the coupling of the pion to the temporal component of the axial current)
and the quark condensate is given in 
 Sec.~\ref{sec.180903.1}. An experimental important source of information for constraining the different approaches to 
nuclear physics is the spectrum of pionic atoms. Although the latter tend to constrain mostly the region of 
 low densities in the nucleus, and it is not very sensitive to in-medium $NN$ interactions (as we explain below),
it is also true that a controlled extrapolation (whenever is possible) of sound theories to higher densities could be  
 of importance for the study of neutron stars. Some concluding remarks are gathered in Sec.~\ref{sec:conc}.

\section{Chiral Power Counting}
\label{sec:pw}

In Ref.\cite{prcoller} the effective chiral Lagrangian for pions is determined in the nuclear medium
 in presence of external sources. 
For that the Fermi seas of protons and neutrons are integrated out making use of functional techniques. 
A similar approach is followed in Ref.\cite{sainio}  for the case of only one nucleon. 
If we write a general chiral Lagrangian density in terms of an increasing number of baryon fields $\psi(x)$ as
\begin{align}
{\cal L}_\chi={\cal L}_{\pi\pi}+{\cal L}_{\bar{\psi}\psi}+{\cal L}_{\bar{\psi}\bar{\psi}\psi\psi}+\ldots
\label{lagnn}
\end{align}
Ref.\cite{prcoller} only retains ${\cal L}_{\pi\pi}$ and ${\cal L}_{\bar{\psi}\psi}$,
with the latter written as
\begin{align}
  \label{181228.1}
  {\cal L}_{\bar{\psi}\psi}&=\bar{\psi}(x)D(x)\psi(x)~,
\end{align}
in terms of the differential operator $D(x)$. This operator is  decomposed as
$D(x)=D_0(x)-A(x)$,
where $D_0(x)=i\gamma^\mu \partial_\mu -m $ is the Dirac operator for the free motion of the nucleons and
$-A(x)$ is the interacting part. In the derivations of Ref.~\cite{annp}, $D^{-1}$ is expressed as
\begin{align}
  \label{181228.2}
  D^{-1}=\left[I-D_0^{-1}A\right]^{-1}D_0^{-1}~.
  \end{align}

Also  Ref.\cite{prcoller} established the concept of an in-medium generalized vertex (IGV).
The latter results by connecting several vacuum operators $\Gamma$, defined as 
\begin{align}
  \label{181228.3}
\Gamma&=-iA\left[I-D_0^{-1}A\right]^{-1}~,
\end{align}
by real nucleons in the Fermi sea.  Notice that the vacuum operator $\Gamma$ is non-local in general, as it arises by
connecting vacuum interacting local vertices $A(x)$ through the free baryon propagator $D_0^{-1}$. Namely, we have
the geometric series expansion for $\Gamma$ as, $\Gamma=A+AD_0^{-1}A+\ldots$ cf. Fig.~\ref{fig.181228.1}. 
An IGV is schematically shown in Fig.~\ref{fig:mgv} where the thick arc segment indicates the insertion of a Fermi sea.
At least one is necessary because otherwise we would have a vacuum closed nucleon loop
which  is not explicitly taken into account in a low-energy effective field theory.\footnote{Its effects are included in the
  renormalization of the vacuum ChPT calculations.}
On the other hand, the filled larger circles in Fig.~\ref{fig:mgv} 
indicate a bilinear baryon vertex from ${\cal L}_{\bar{\psi}\psi}$ ,$-iA$, while the dots refer to the insertion of any number of them.
In this figure the thin segments correspond to fermionic propagators, either to vacuum ones, $i D_0^{-1}$, or to Fermi-sea
insertions. Within the relativistic formalism of Ref.~\cite{prcoller} we can state the following Feynman rules for evaluating the in-medium 
diagrams. For an IGV with $n$ Fermi-sea insertions we have the combinatoric factor $(-1)^{n+1}/n$, and for each 
 Fermi sea there is the factor $(2\pi)\delta(p^2-m^2)\theta(p^0)(p\!\!\!/+m)\theta(\xi-|\vp|)$, where $p$ is the four-momentum 
 flowing through the thick line, $\xi$ is a Fermi momentum and $\theta(x)$ is the standard Heaviside function. 
A factor $-iA$ accompanies any vacuum vertex and for the vacuum propagators of baryons we have 
$iD_0^{-1}=i(p\!\!\!/+m)/(p^2-m^2+i\epsilon)$, $\epsilon\to 0^+$. In addition, one also could have pion lines as external or 
 internal particles connecting different bilinear vertices. In this respect an IGV behaves analogously to a local vertex in 
 ${\cal L}_{\pi\pi}$. 

\begin{figure}
	\begin{center}
		\begin{axopicture}(300,60)(0,30)
			\SetWidth{1.5}
			\ECirc(25,50){10}
			\SetWidth{0.5}
			\Text(25,50){$\Gamma$}
			\Text(45,50){$=$}
			\GCirc(65,50){10}{.1}
			\Text(85,50){$+$}
			\GCirc(105,50){10}{0.1}
			\Line(135,50)(115,50)
			\Text(125,50){{\small $/$}}
			\GCirc(145,50){10}{0.1}
			\Text(165,50){{$+$}}
			\GCirc(185,50){10}{0.1}
			\Line(215,50)(195,50)
			\Text(205,50){{\small $/$}}
			\GCirc(225,50){10}{0.1}
			\Line(255,50)(235,50)
			\Text(245,50){{\small $/$}}
			\GCirc(265,50){10}{0.1}
			\Text(285,50){$+$}
			\Text(304,50){$\ldots$}
		\end{axopicture}
		\caption[pilf]{\protect \small
Expansion of the non-local vacuum vertex $\Gamma$.
Every solid line with the slash corresponds to a vacuum baryon propagator and
each circle to the insertion of an operator $-iA$ from  ̄$\bar{\psi}D\psi$.
			\label{fig.181228.1}}
	\end{center}
\end{figure}

Based on these results Ref.\cite{annp} derived a chiral power counting in the nuclear medium for pion-mediated interactions.
In this reference ${\cal L}_{\bar{\psi}\psi}$ is replaced by ${\cal L}_{\pi N}$, which is
the bilinear Lagrangian in the nucleon fields containing pions and external sources. 
For illustration we plot in Fig.~\ref{fig.181228.2} several Feynman diagrams involving different number of IGVs 
and Fermi-sea insertions in each of them. From left to right, the first two diagrams are of Hartree type and the last one is a Fock 
diagram. They are possible diagrams in the evaluation of the energy density in nuclear matter, cf. Sec.~\ref{sec.180829.1}. 
The first two diagrams contain two IGVs, and the last diagram has only one IGV. The solid line in the second diagram with a slash is a 
free baryon propagator, while the rest of nucleon lines are Fermi-sea insertions. The dashed lines correspond to pion exchanges. 
More examples of in-medium diagrams are discussed below in the actual applications considered explicitly. 

\begin{figure}[ht]
	\psfrag{q1}{$q$}
	\psfrag{l}{$i$}
	\psfrag{m}{$j$}
	\centerline{\epsfig{file=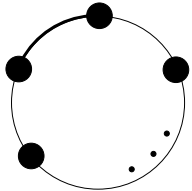,width=0.14 \textwidth,angle=0}}
	\vspace{0.2cm}
	\caption[pilf]{\protect \small
		In-medium generalized vertex (IGV). 
		The thick solid line corresponds to a Fermi-sea insertion, 
the thin ones are either vacuum baryon propagators or extra Fermi-sea insertions, 
while the filled circles are bilinear baryon vertices $-iA$ from ${\cal L}_{\bar{\psi}\psi}$.
		\label{fig:mgv}}
\end{figure} 

\begin{figure}
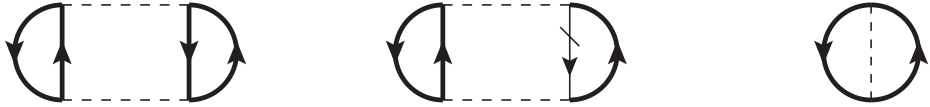

	\begin{center}
		\SetScale{1.2}
		\begin{axopicture}(300,60)(0,15)
		\SetWidth{1.5}
		  \Line[arrow](25,35)(25,65)    
		  \Arc[arrow](25,50)(15,90,270) 
		  \Line[arrow](65,65)(65,35)    
          \Arc[arrow](65,50)(15,270,90) 
		  \Line[arrow](145,35)(145,65)    
		  \Arc[arrow](145,50)(15,90,270) 
		  \Arc[arrow](185,50)(15,270,90) 
\SetWidth{0.5}
          \Line[dash](25,65)(65,65)
          \Line[dash](25,35)(65,35)
          \Line[dash](145,65)(185,65)
          \Line[dash](145,35)(185,35)
		  \Line[arrow,arrowpos=0.65](185,65)(185,35)    
	\Line(182,58)(188,52)
		\SetWidth{1.5}
		\Arc[arrow](280,50)(15,90,270)    
		\Arc[arrow](280,50)(15,270,90) 
		\SetWidth{0.5}
		\Line[dash](280,35)(280,65) 
		\end{axopicture}
\caption[pilf]{\protect \small
From left to right, we have two possible Hartree diagrams and  
one possible Fock diagram for the calculation of the energy density in nuclear matter. 
The first and second diagrams contain two IGVs while the last one has only one IGV. 
The two IGVs in the first diagram have two Fermi-sea insertions, while the last IGV in the second diagram has only one 
Fermi-sea insertion. In the third diagram all the baryon lines are Fermi-sea insertions. 
 The dashed lines correspond to pions. For simplifying the diagrams we have not drawn the bilinear vertices as filled circles.
			\label{fig.181228.2}}
	\end{center}
\end{figure}

In relation with the infrared enhancements of nucleon propagators discussed by Weinberg \cite{wein1,wein2} for the vacuum 
propagation of two o more nucleon states, 
 Ref.\cite{annp} stressed that within a nuclear environment even a single nucleon propagator could have such 
 enhancements, driving to the so-called non-standard chiral counting. 
 To see this, note that a soft momentum $Q$, related to pions or external sources, 
can be associated to any of the bilinear vertices. 
This, together with the Dirac delta function of four-momentum conservation, implies that the momenta running along 
the nucleon propagators in Fig.~\ref{fig:mgv} just differ from each other by quantities of ${\cal O}(Q)$. 
Denoting by $k$ the on-shell four-momenta associated with one Fermi-sea insertion in the in-medium generalized vertex, 
the four-momentum running through the $j_{th}$ nucleon propagator can be written as $p_j=k+Q_j$. In this way,
\begin{align}
iD_0^{-1}(p_j)&=i\frac{\barr{k}+\barr{Q}_j+m}{(k+Q_j)^2-m^2+i\epsilon}=
 i \frac{\barr{k}+\barr{Q}_j+m}{Q_j^2+2 Q_j^0 E(\vk) -2 {\mathbf{Q}}_j
 \mathbf{k}+i\epsilon}~,
\label{pro.1}
\end{align}
 and $E(\vk)=\vk^2/2m$.

Two different situations occur depending on the value of $Q_j^0$. If $Q_j^0={\cal O}(p)$  
one has the standard counting so that the chiral expansion of the propagator in Eq.~\eqref{pro.1} is
\begin{align}
 iD_0^{-1}(p_j)&=i \frac{\barr{k}+\barr{Q}_j+m}{2 Q^0_j m+i\epsilon} 
 \left(1-\frac{Q_j^2-2 {\mathbf{Q}}_j \cdot \mathbf{k}}{2Q_j^0 m}+\Opd \right)~.
\end{align}
Thus, $iD_0^{-1}$ counts as a quantity of ${\cal O}(p^{-1})$. 
But it could also occur that $Q_j^0$ is ${\cal O}(E(\vk))$, that is, of the order of a kinetic nucleon energy 
in the nuclear medium or even lower, indeed it might even vanish \cite{annp}. 
The dominant term in Eq.~\eqref{pro.1} is then
\begin{align}
iD_0^{-1}&=-i\frac{\barr{k}+\barr{Q}_j+m}{\mathbf{Q}^2_j+2\mathbf{Q}_j\cdot\vk-i\epsilon}~,
\end{align}
and the nucleon propagator should be counted as ${\cal O}(p^{-2})$, instead of the previous ${\cal O}(p^{-1})$. 
This is referred as the non-standard case in Ref.\cite{annp}. 
It is well known that this situation occurs already in vacuum when considering the two-nucleon 
reducible diagrams in $NN$ scattering. 
 To face this problem Ref.\cite{wein1} advocates for solving non-perturbatively a Lippmann-Schwinger equation 
with the $NN$ potential given by the two-nucleon irreducible diagrams. 
The case of nucleon reducible diagrams also occurs in the nuclear medium where there 
are an infinite number of nucleons provided by the Fermi seas.

The Ref.~\cite{lacour.180827.2} extended the results of Refs.\cite{prcoller,annp} in a twofold way.
i) Chiral Lagrangians with an arbitrary number of baryon fields (bilinear, quartic, etc) were considered.
First only bilinear vertices like in Refs.\cite{prcoller,annp} are taken into account, 
but the  exchanges of heavy meson fields of any type are allowed in addition to pion exchanges. 
The former should be considered as merely auxiliary fields that allow one to find a tractable
representation of the multi-nucleon interactions that result when the masses of the heavy mesons tend to infinity. 
This is similar to the Hubbard-Stratonovich transformation.
ii) The non-standard counting is taken from the start and any nucleon propagator is counted as ${\cal O}(p^{-2})$.
In this way, no diagram, whose chiral order is actually lower than expected if the nucleon propagators were counted assuming 
the standard rules, is lost. This was a step forward in the literature.


The in-medium chiral  power counting deduced in Ref.~\cite{lacour.180827.2} proceeds in the following manner. 
Let us denote by $H$ the heavy mesons responsible, because of their exchanges between bilinear vertices in the limit of infinite mass,  
of the contact multi-nucleon interactions, $NN$, $NNN$, etc. 
From the counting point of view there is a clear similarity between the interactions driven by the exchanges of $H$ 
and $\pi$ fields as both emerge from bilinear vertices. 
The large mass of the former is responsible of the local character of the induced interactions. 
A heavy meson propagator is counted as $p^0$.

The chiral order of a given diagram is represented by $\nu$ and it is given by
\be
\nu=4L_H+4L_\pi-2I_\pi+\sum_{i=1}^{V_\rho}\left[\sum_{j}d_j-2m_i\right]+\sum_{i=1}^{V_\pi}\delta_i
+\sum_{i=1}^{V_\rho}3~.
\label{count}
\ee
Here, $V_\rho$ is the number of IGVs, $m_i$ is the number of nucleon propagators   in the $i_{th}$ IGV minus one, 
which is the one corresponding to the needed  Fermi-sea insertion for every in-medium generalized vertex (and that gives rise 
to the last term in the previous equation).  
In addition, $d_i$ is the chiral order of the $i_{th}$ vertex bilinear in the baryon fields, 
$\delta_i$ is the chiral order of a vertex without baryons (only pions and external sources) 
and $V_\pi$ is the number of the latter ones. 
As usual, $L_\pi$ is the number of pion loops and $I_\pi$ is the number of internal pion lines. 
 Finally, $L_H$ is the number of loops due to  the internal $H$ lines.
 
 Let us note that associated with the bilinear vertices in an IGV one has  four-momentum conservation 
delta functions that can be used to fix the momentum of each of the baryon lines joining them, 
except one for the running momentum due to the Fermi-sea insertion.
Of course, this cannot be fixed because one four-momentum delta function has to do with the conservation of the total four-momentum. 
This is the reason why we referred above only to loops  attached to meson lines and not to  baryon ones.
We then have the following expression for the total number of loops associated with running meson momenta, 
\begin{align}
\label{181229.1}
L_\pi+L_H=I_\pi+I_H-V_\rho-V_\pi+1~.
\end{align}
There is also a well-known expression relating the total number of meson lines (both internal as external) with 
the number of meson lines attached to the different vertices. It reads,
\be
2I_H+2I_\pi+E_\pi=\sum_{i=1}^V v_i+\sum_{i=1}^{V_\pi}n_i~,
\label{lines}
\ee
where $V$ is the total number of bilinear vertices, $v_i$ is the number of meson lines attached to the 
$i_{th}$ bilinear vertex, $n_i$ is the number of pions in the $i_{th}$ meson vertex and $E_\pi$ is the 
number of external pion lines. Taking into account Eq.~\eqref{181229.1} one has for the first three terms of Eq.~\eqref{count},
\be
4L_H+4L_\pi-2I_\pi=4I_H+2I_\pi-4V_\rho-4V_\pi+4~.
\label{eqint}
\ee
Now considering Eq.~\eqref{lines}  in the right hand side of this equation, it becomes
\be
4L_H+4L_\pi-2I_\pi=2I_H-E_\pi+\sum_{i=1}^V v_i+\sum_{i=1}^{V_\pi}n_i-4V_\rho-4 V_\pi+4~.
\ee

Substituting the previous line in Eq.~\eqref{count},
\be
\nu=2I_H-E_\pi+4-4V_\pi+\sum_{i=1}^{V_\pi}(\delta_i+n_i)+\sum_{i=1}^V (d_i+v_i)-2 \varkappa - V_\rho~.
\ee
with $\varkappa=\sum_{i=1}^{V_\rho}m_i$. 
We now employ that $V_\rho+\varkappa=V$, and $2 I_H=\sum_{i=1}^V \omega_i$, 
where $\omega_i$ is the number of heavy meson internal lines for the $i_{th}$ bilinear vertex.
Then, we arrive to our final equations,
\begin{align}
\label{fff}
\nu&=4-E_\pi+\sum_{i=1}^{V_\pi}(\delta_i+n_i-4)+\sum_{i=1}^V(d_i+v_i+\omega_i-2)+ V_\rho~.
\end{align}

Note the very important property that $\nu$  is bounded from below for any Feynman diagram with a given number of 
external pion lines and sources. For the first sum in the right hand side of Eq.~\eqref{fff}  one has that 
\be
\label{181229.2}
\delta_i+n_i-4\geq 0~,
\ee
as $\delta_i= 2 n$ ($n\in \mathbb{N}\backslash \{0\}$) and $n_i\geq 2$, except possibly for a finite number of vertices in a  
Feynman diagram that could contain less than two pion 
lines with $\delta_i=2$. Similarly 
\be
\label{181229.3}
d_i+\omega_i+v_i-2\geq 0~,
\ee
since for pion-nucleon vertices  $d_i\geq 1$ and $v_i\geq 1$, except for the finite 
number of vertices in a given Feynman diagram which would not have pion lines but only external sources 
from ${\cal L}_{\pi N}$ with $d_i=1$. 
For the bilinear vertices mediated by heavy lines this is also the case because 
 $\omega_i\geq 1$, $v_i\ge \omega_i$, though here $d_i\geq 0$.  
Let us stress that the vertices that would not fulfill Eqs.~\eqref{181229.2} and \eqref{181229.3} are finite in number because 
one always has external sources attached to them, whose number is finite. For the evaluation of the correlation function of $n$ 
external sources they could contribute at most a negative 
chiral power of $-2n$, taking the most disfavorable case in which every external source is attached to a different pure mesonic vertex 
with $\delta_i=2$ and $n_i=0$.
 It is specially important to note that adding a new IGV to a connected diagram 
increases the counting at least by one unit because of the last term $V_\rho$ in Eq.~\eqref{fff}.

The number $ \nu$ given in Eq.~\eqref{fff} represents a lower bound for the actual chiral power of a
diagram, let us call this by $\mu$, and then $\mu\geq \nu$. 
The reason why $\mu$ might be different from $\nu$ is because 
the nucleon propagators are counted always as ${\cal O}(p^{-2})$, while for some diagrams there could be baryon propagators 
following the standard counting. 
The  point of Eq.~\eqref{fff} is that it allows to ensure that no other
contributions to those already considered would have a lower chiral order. 
As a result, it can handle systematically the anomalous chiral counting introduce above. 

From Eq.~\eqref{fff} it is clear that one can augment the number of lines in a diagram without increasing the power counting 
by:
\begin{enumerate}
\item Adding pion lines attached to the lowest-order meson vertices, $\delta_i=n_i=2$.
\item Adding pion lines attached to the lowest-order meson-baryon vertices, $d_i=v_i=1$.
\item Adding heavy meson lines attached to the lowest-order bilinear vertices, $d_i=0$, $w_i=1$.
\end{enumerate}

There is no way to decrease the chiral order for a given process. 
We apply Eq.~\eqref{fff}  by increasing step by step $V_\rho$ up to the order pursued.
For each $V_\rho$ then we look for those diagrams that do not further increase the order according to the previous list.
Some of these diagrams are indeed of higher order and one can refrain from calculating them by establishing which of the baryon propagators  scale as ${\cal O}(p^{-1})$. 
In this way, the  actual chiral order of the diagrams is determined and one can select those diagrams that correspond to the precision required.

Let us now elaborate on the important scales appearing in the low-momentum calculations in nuclear matter. 
Taking into account that a term in a Lagrangian density has dimensions $[p^4]$, we have that the pure meson vertices 
with $a$ pion fields $\phi$ and $b$ derivatives (every of them gives rise to a soft 
momentum factor of order $p$), contributes with a monomial that scales as
\be
\label{dan2}
\frac{\phi^a}{f^{a-2}}\frac{p^b}{\Lambda^{b-2}}~.
\ee
where $f$ is the pion weak decay constant  in the chiral limit
 and $\Lambda$ is a generic hard scale that does not vanish in the chiral limit. 
In addition, we also have the IGVs that behave as effective vertices connected 
by meson lines. Each of them contains at least one three-momentum integration up to a Fermi momentum of at least one Fermi-sea insertion, 
so that instead of Eq.~\eqref{dan2} we have contributions that scale typically as
\be
\label{dan3}
\rho\frac{\phi^a}{f^{a}}\frac{p^b}{\Lambda^{b-1}}~,
\ee
with $\rho$ the nuclear matter density. 
One expects a suppression of a chiral power 3 for the in-medium contributions because of the $\rho$ factor. 
Indeed, by directly taking the quotient of Eqs.~\eqref{dan3} and \eqref{dan2} we have
\begin{align}
\label{181229.4}
\frac{\rho}{f^2 \Lambda}={\cal O}\left( \frac{p^3}{\Lambda(\upsilon \pi f)^2}\right)~,
\end{align}
with $\upsilon$ a number of ${\cal O}(1)$. 
This equation implies that compared with the vacuum scale the in-medium one is expected to be somewhat reduced by the numerical factors from the expression of $\rho$. 
Taking literally the left hand side of Eq.~\eqref{181229.4} (although ${\cal O}(1)$ numerical factors could float), one has that while
 in vacuum the loop scale is $4\pi f$ (see e.g. chapter 5.9 of Ref.~\cite{georgi.181229.1}),  it would be $3\pi f$ for pure neutron matter and $3\pi f/2$ for symmetric nuclear matter. We have taken into account that $\rho=\xi^3/3\pi^2$ in the former case 
and $\rho=2\xi^3/3\pi^2$ in the latter. 
Thus, in the nuclear medium the expansion scale $\Lambda_\xi$ is expected to be given by $\Lambda_\xi=\upsilon \pi f$, and it is not really larger than the mass of the $\rho(770)$ resonance, as $4\pi f$ is. In this way, 
for symmetric nuclear matter at the saturation density with $\xi\simeq 2m_\pi$ we expect corrections to our results 
of ${\cal O}(2m_\pi/\Lambda_\xi)\simeq 0.3$.
   
The presence of the anomalous counting makes that instead of $p^{-1}$ associated with a baryon propagator in the IGV 
one has $(p^2/2m)^{-1}$. 
To fix ideas, let us assume that we have one of such enhanced baryon propagators. Then, instead of Eq.~\eqref{dan3} we would have
\be
\label{dan3b}
\rho\frac{\phi^a}{f^{a}}\frac{2m p^{b-1}}{\Lambda^{b-1}}~.
\ee
Taking the quotient of this equation with Eq.~\eqref{dan2} for the vacuum contributions, we 
have now the suppression factor
\begin{align}
  \label{dan3c}
\frac{2m \rho}{f^2\Lambda p}\sim \frac{2m \xi}{(\upsilon \pi f)^2}\frac{\xi}{\Lambda}~.
\end{align}
In this equation the last term on the right hand side is the same one as in the standard case, 
but now the in-medium scale in the first factor is much reduced. Let us denote the latter by $\Lambda_{\rm low}$, 
which is then given by
\begin{align}
\label{dan3d}
\Lambda_{\rm low}\simeq \frac{(\upsilon \pi f)^2}{2m}\ll \upsilon \pi f~.
\end{align}
In many instances the factor $2m$ in the denominator is multiplied by the nucleon axial coupling squared, $g_A^2\simeq 1.7$, from 
the one pion exchange. 
Therefore, one has that $\Lambda_{\rm low}$ compared to $\upsilon \pi f$ is reduced typically by a factor around 2--4. 

Due to the smallness of $\Lambda_{\rm low}$ it is essential for systematic calculations in the nuclear medium to derive an in-medium power counting. 
This is the most important point of the power counting developed in Ref.~\cite{lacour.180827.2} and summarized here, driving to 
Eq.~\eqref{fff}. 
Despite the importance of this fact our research is the only one within the many groups applying ChPT to the nuclear medium that proceeds strictly according to an in-medium power counting that takes into account the infrared enhancements of the baryon propagators. 
Typically, the approaches followed in the literature, e.g. Refs.~\cite{fiorilla.181229.1,holt.181229.1,sch.180902.1} and references therein, are based on a straightforward application of the {\it vacuum} chiral power counting to in-medium calculations. 
The goal in those calculations is to apply perturbation theory to determine different magnitudes or to calculate input quantities that are used within more or less sophisticated many-body methods. 

\section{Nucleon-nucleon interactions}
\label{sec:nn-int}

The inclusion of the $NN$ interactions for the calculation of different nuclear-matter properties takes place at NLO, 
because they require  at least $V_\rho=2$. 
As a result,  for the NLO calculations it is only necessary to work them out at the lowest chiral order or  ${\cal O}(p^0)$.    
The $NN$ interactions were approximated in Refs.~\cite{lacour.180827.3,lacour.180827.4} 
by employing a handy approach based on Unitary Chiral Perturbation Theory (UChPT), whose main facets are reviewed here 
\cite{nn,lacour.180827.3}. 
 Of course, it is desirable, as future advances in the field, to include the $NN$ interactions in the powerful novel  
non-perturbative theory based on the application of the exact calculation of the discontinuity along the 
left-hand cut (LHC) of a $NN$ partial-wave amplitude and then applying the exact $N/D$ method \cite{entem.180828.1}.

\subsection{Vacuum $NN$ interactions}
\label{sec:fnn}

For the two-nucleon irreducible diagrams we follow the standard chiral counting \cite{wein1,wein2}. Accordingly,  
 the lowest order amplitudes, which count as ${\cal O}(p^0$), are given by the quartic nucleon Lagrangian,  
without quark masses or derivatives, and by the one-pion exchange (OPE) expressed in terms of the 
lowest order pion-nucleon coupling $\frac{i g_A}{2f}\vec{\sigma}\cdot \vq \, \vec{\tau}\cdot\vec{\pi}$. 
Here $\sigma_i$ and $\tau_i$ are the Pauli matrices in the spin and isospin spaces, in this order. 
The ${\cal O}(p^0)$ lowest order four nucleon Lagrangian  is \cite{wein2} 
\be
{\cal L}_{NN}^{(0)}=-\frac{1}{2}C_S (\overline{N}N)(\overline{N}N)
-\frac{1}{2}C_T(\overline{N}\vec{\sigma} N)(\overline{N}\vec{\sigma} N)~,
\label{lnn}
\ee
which only contributes to the  S-wave nucleon-nucleon scattering. 
The resulting scattering amplitude from the previous Lagrangian for the  process 
$N_{s_1,i_1}(\vp_1) N_{s_2,i_2}(\vp_2) \to  N_{s_3,i_3}(\vp_3) N_{s_4,i_4}(\vp_4)$, 
where $s_m$ and $i_m$ are spin and isospin labels,  is
\begin{align}
T_{NN}^{c}&=-C_S\left(\delta_{s_3 s_1}\delta_{s_4 s_2}\,\delta_{i_3i_1} \delta_{i_4 i_2}-\delta_{s_3 s_2}\delta_{s_4 s_1}\,\delta_{i_3i_2} \delta_{i_4i_1}\right)\nn\\
&-C_T\left(\vec{\sigma}_{s_3 s_1}\cdot \vec{\sigma}_{s_4 s_2}\, \delta_{i_3i_1}\delta_{i_4i_2}-\vec{\sigma}_{s_3 s_2}\cdot \vec{\sigma}_{s_4 s_1} \,\delta_{i_3i_2}\delta_{i_4i_1}\right)~.
\label{feynman}
\end{align}
Because of the selection rule $S+\ell+I=$odd (with $S$ and $I$ the total spin and isospin of the system and $\ell$ 
its orbital angular momentum), that holds for any possible $NN$ partial wave due to the Fermi statistics, 
the only partial waves from Eq.~\eqref{feynman} are
\begin{align}
T_{NN}^c(^1S_0)&=-2(C_S-3C_T)~,\nn\\
T_{NN}^c(^3S_1)&=-2(C_S+C_T)~.
\label{a.local}
\end{align}

\begin{figure}[ht]
\psfrag{k}{$k$}
\psfrag{p}{$p$}
\psfrag{l}{$\ell$}
\psfrag{pi}{$\pi$}
\psfrag{r}{$k-\ell$}
\centerline{\epsfig{file=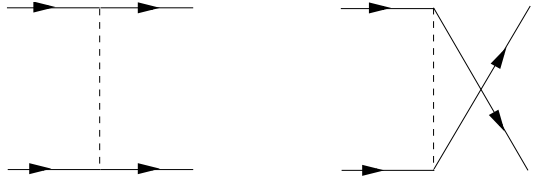,width=.4\textwidth,angle=0}}
\vspace{0.2cm}
\caption[pilf]{\protect \small
One-pion exchange diagrams for the nucleon-nucleon scattering amplitude. 
The digram on the left corresponds to the direct contribution and  the one on the right to the exchange amplitude.
\label{fig:1pi}}
\end{figure} 
In addition, one also has the OPE amplitudes, depicted in Fig.~\ref{fig:1pi}, which are given by
\begin{align}
T_{NN}^{1\pi} = \frac{g_A^2}{4f^2}\left[
\frac{
(\vec{\tau}_{i_3i_1}\cdot \vec{\tau}_{i_4i_2})
(\vec{\sigma}\cdot \vq)_{s_3s_1}(\vec{\sigma}\cdot \vq)_{s_4s_2}}{\vq^2+m_\pi^2-i\epsilon}
- \frac{(\vec{\tau}_{i_4i_1}\cdot 
\vec{\tau}_{i_3i_2})(\vec{\sigma}\cdot \vq')_{s_4s_1}(\vec{\sigma}\cdot \vq')_{s_3s_2}}{{\vq'}^2+m_\pi^2-i\epsilon}
\right]~,
\label{1pi.gen}
\end{align}
with $\vq=\vp_3-\vp_1$ and $\vq'=\vp_4-\vp_1$. For the singlet case ($S=0$)  and $I=0,~1$ one has,
\begin{align}
T_{NN}^{1\pi}(S=0,I=0)&=\frac{3g_A^2}{4f^2}\left[
\frac{\vq^2}{\vq^2+m_\pi^2-i\epsilon}-\frac{{\vq'}^2}{{\vq'}^2+m_\pi^2-i\epsilon}
\right]~,\nn\\
T_{NN}^{1\pi}(S=0,I=1)&=\frac{-g_A^2}{4f^2}\left[\frac{\vq^2}{\vq^2+m_\pi^2-i\epsilon}
+\frac{{\vq'}^2}{{\vq'}^2+m_\pi^2-i\epsilon}\right]~.
\label{1pi.sin}
\end{align}
 For the triplet case ($S=1$) a $3\times 3$ matrix results with labels given by the third component of the total spin, 
 $\sigma_f$, $\sigma_i$, with the subscripts $f$ (rows) and $i$ (columns) 
 referring to the final and initial third components, respectively:
\begin{equation}
||B_{\sigma_f \sigma_i}|| = \left(
\begin{array}{l|lll}
&-1 & 0 & +1 \\
\hline
-1 & q_3^2 & -\sqrt{2}(q_1+iq_2)q_3 & (q_1+iq_2)^2\\
0 & -\sqrt{2}(q_1-iq_2)q_3 & q_1^2+q_2^2-q_3^2 & \sqrt{2}(q_1+iq_2)q_3\\
+1& (q_1-iq_2)^2& \sqrt{2}(q_1-iq_2)q_3 & q_3^2 
\end{array}
\right)
\label{180828.1}
\end{equation}
The Cartesian coordinates of $\vq$ are indicated as subscripts. 
 The isoscalar and isovector amplitudes are, in this order, 
\begin{align}
T^{1\pi}_{NN;\sigma_f \sigma_i}(S=1,I=0) &= \frac{-3g_A^2}{4f^2} \frac{B_{\sigma_f \sigma_i}}{\vq^2+m_\pi^2-i\epsilon} + (\vq\leftrightarrow \vq')~.\nn\\
T^{1\pi}_{NN;\sigma_f \sigma_i}(S=1,I=1) &= \frac{g_A^2}{4f^2} \frac{B_{\sigma_f \sigma_i}}{\vq^2+m_\pi^2-i\epsilon} -(\vq\leftrightarrow \vq')~.
\label{1pi.s1t1}
\end{align}
Considering the Eqs.~\eqref{1pi.sin} and \eqref{1pi.s1t1},  
one can calculate the corresponding $NN$ OPE partial-wave amplitudes. 
Since the OPE scattering  amplitudes in these equations are already given 
in terms of $NN$ states with definite spin and isospin, this calculation simplifies to 
\begin{align}
N_{JI}^{1\pi}(\bar{\ell},\ell,S)&=\frac{Y_\ell^0(\hat{\vz})}{2J+1}\sum_{\sigma_i,\sigma_f=-S}^{S}
(0\sigma_i \sigma_i|\ell SJ)(\bar{m}\sigma_f \sigma_i|\bar{\ell}SJ) \int d\hat{\vp}
\,T^{1\pi}_{NN;\sigma_f\sigma_i}(S,I)Y_{\bar{\ell}}^{\bar{m}}(\hat{\vp})^*~,
\label{1pi.pw}
\end{align}
where we use the notation $(m_1 m_2 m_3|j_1 j_2 j_3)$ for the Clebsch-Gordan coefficients of the composition of two 
angular momenta $j_1$ and $j_2$ to give $j_3$, with third components $m_i$, and the $Y_{\ell}^m(\vp)$ are the spherical harmonics. 
In practical evaluations we keep all the partial waves up to and including $\ell=3$. 
Explicit expressions for the resulting OPE $NN$ partial waves $N_{JI}^{1\pi}(\bar{\ell},\ell,S)$ 
are given in Appendix \ref{app:1pi}.

\begin{figure}[ht]
\psfrag{k}{$k$}
\psfrag{p}{$p$}
\psfrag{l}{$\ell$}
\psfrag{pi}{$\pi$}
\psfrag{r}{$k-\ell$}
\centerline{\epsfig{file=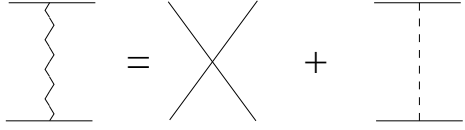,width=.35\textwidth,angle=0}}
\vspace{0.2cm}
\caption[pilf]{\protect \small
The exchange of a wiggly line between two nucleons correspond to the sum of the local and OPE contributions.
\label{fig:wig}}
\end{figure}

The sum of the local contributions, Eq.~\eqref{a.local}, and the OPE partial-wave amplitudes, Eq.~\eqref{1pi.pw},  
is represented diagrammatically in the following by the exchange of a wiggly line as in Fig.~\ref{fig:wig}.

\begin{figure}[ht]
\psfrag{k}{$k$}
\psfrag{p}{$p$}
\psfrag{l}{$\ell$}
\psfrag{pi}{$\pi$}
\psfrag{r}{$k-\ell$}
\centerline{\epsfig{file=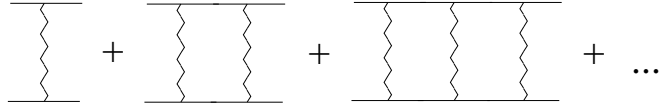,width=.5\textwidth,angle=0}}
\vspace{0.2cm}
\caption[pilf]{\protect \small
Resummation of the two-nucleon reducible diagrams. 
This is also referred  as the resummation of the right hand cut or unitarity cut.
\label{fig:sum}}
\end{figure} 

 As it is well known, Weinberg argued \cite{wein1,wein2} that due to the large nucleon mass one has to 
 resum the two-nucleon reducible diagrams, 
 as it is schematically depicted in Fig.~\ref{fig:sum}. 
Indeed, this is also required by applying the power counting of Eq.~\eqref{fff}, cf. points 2  and 3 for increasing the number of 
internal lines in a Feynman graph without increasing the chiral order.
 For these diagrams  the nucleon propagators follow the non-standard counting and each of them is ${\cal O}(p^{-2})$. 
The two nucleon propagators altogether are ${\cal O}(p^{-4})$, which are multiplied by the ${\cal O}(p^4)$ contribution from 
the measure of the loop integral, and then an ${\cal O}(p^0)$ contribution results. 
Thus, it does not rise the chiral order when the wiggly lines are evaluated at lowest order, ${\cal O}(p^0)$, 
 and the series of diagrams in Fig.~\ref{fig:sum} must be resummed. 
The resummation of the two-nucleon reducible diagrams makes the resulting amplitude to fulfill unitarity. 
For this resummation, Refs.~\cite{lacour.180827.2,lacour.180827.3} follow the techniques of UChPT \cite{nd,meis,npa,report1},  
which allow to resum the right hand cut (RHC) or unitarity cut  partial wave by partial wave. 
UChPT has also been applied with great success in meson-meson \cite{nd,report1,alba1,alba2,guoelvira,kang} and
meson-baryon scattering \cite{ramos,meis,prl1,prl2,ojpa,guoremake}, among many other references. For the study of the  
$NN$ interactions UChPT was already applied in Ref.~\cite{nn}.

\begin{figure}
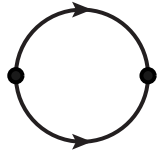

\begin{center}
\begin{axopicture}(200,100)(0,10)

\SetWidth{1.5}
\Arc[arrow,clockwise](100,50)(25,175,5) 
\Arc[arrow](100,50)(25,185,355) 
\GCirc(75,50){2.5}{.1}
\GCirc(125,50){2.5}{.1}
\end{axopicture}
\caption[pilf]{\protect \small
Unitarity loop corresponding to the function $g(p^2)$, cf. Eqs.~\eqref{int.g} and \eqref{181231.1}.
\label{fig:g}}
\end{center} 
\end{figure}

The master equation for UChPT is the same independently of whether we have fermions, 
mesons or both in the scattering process and can be written as \cite{meis}
\be
T_{JI}(\bar{\ell},\ell,S)=\left[I + N_{JI}(\bar{\ell},\ell,S) \cdot g(p^2)\right]^{-1}\cdot N_{JI}(\bar{\ell},\ell,S)~,
\label{master}
\ee
where we have used a matrix notation valid also for coupled channels mixing different orbital angular momenta, 
$\bar{\ell}$ and $\ell$. In the previous equation $g(p^2)$ is the unitarity loop function drawn in Fig.~\ref{fig:g}. 
For $NN$ scattering with non-relativistic kinematics the function 
$g(p^2)$ reads
\begin{align}
\label{181231.1}
g(p^2)&=g_0-i\frac{m \sqrt{p^2}}{4\pi}~,
\end{align}
with $g_0$  the value of $g(p^2)$ at threshold, $p^2=0$.  
In the case of coupled partial waves, $g(p^2)$ is actually a diagonal matrix that corresponds to the product of the right hand side of Eq.~\eqref{181231.1} with the $2\times 2$ identity matrix.

 The Eq.~\eqref{master} results by performing a once subtracted dispersion relation of the inverse of a partial-wave amplitude.  
 The latter  fulfills, because of unitarity,
\be
\left.\hbox{Im}T_{JI}(\bar{\ell},\ell,S)^{-1}\right|_{\bar{\ell}\ell}=-\frac{m p}{4 \pi}\delta_{\bar{\ell}\ell}~,
\label{unitarity}
\ee
in the CM frame and above the elastic threshold. 
 A dispersion relation along the physical energy axis from threshold up to infinity is written. 
 One subtraction is needed because  $p=\sqrt{2 m E}$, with $E$ the kinetic energy of one nucleon in the CM. 
 As a result of this dispersion relation one ends up with a once-subtracted dispersive representation of $g(p^2)$ which 
 reads 
\be
g(p^2)=g(\kappa^2)-\frac{m(p^2-\kappa^2)}{4\pi^2}
\int_0^\infty dk^2\frac{k}{(k^2-p^2-i\epsilon)(k^2-\kappa^2-i\epsilon)}~,
\label{dis.rel.g}
\ee
where $\kappa^2<0$ and then $g(\kappa^2)$ is real because  there is an imaginary part only above threshold. 
This integral can be done explicitly with the result,
\be
g(p^2)=g(\kappa^2)-\frac{i m}{4\pi}\left( \sqrt{p^2}-i\sqrt{|\kappa^2|}\right)=
g_0-i\frac{m \sqrt{p^2}}{4\pi}~.
\label{def.g}
\ee
This function  corresponds to the divergent integral 
\be
g(p^2)=-m \int \frac{d^3 k}{(2\pi)^3}\frac{1}{k^2-p^2-i\epsilon}~.
\label{int.g}
\ee
In terms of a three-momentum cut-off $\Lambda$ the function $g(p^2)$ in the previous equation becomes 
\be
g(p^2)=-\frac{m\Lambda}{2\pi^2}-i\frac{m \sqrt{p^2}}{4\pi}~.
\label{gdc}
\ee
Comparing with Eq.~\eqref{def.g} it follows that 
\be
g_0=-\frac{m\Lambda}{2\pi^2}~.
\label{g0}
\ee
The LHC contributions to $T_{JI}(\bar{\ell},\ell,S)^{-1}$ are parameterized by the inverse of the matrix 
$N_{JI}(\bar{\ell},\ell,S)$, that has only LHC. For clarification, notice that we can write Eq.~\eqref{master} equivalently as 
$T_{JI}(\bar{\ell},\ell,S)=\left[N_{JI}(\bar{\ell},\ell,S)^{-1}+g(p^2)\right]^{-1}$.

Next, $N_{JI}(\bar{\ell},\ell,S)$ is determined by matching with
 the perturbative result in an expansion in powers of $g(p^2)$ of Eq.~\eqref{master}  
 up to the same number of two-nucleon reducible loops (so that the resulting $N_{JI}$ has no LHC by construction). 
The aforementioned expansion corresponds to the geometric series
\begin{align}
T_{JI}(\bar{\ell},\ell,S)&=
N_{JI}(\bar{\ell},\ell,S)
-N_{JI}(\bar{\ell},\ell,S)\cdot g \cdot N_{JI}(\bar{\ell},\ell,S)\nn\\
&+N_{JI}(\bar{\ell},\ell,S)\cdot g\cdot N_{JI}(\bar{\ell},\ell,S)\cdot g \cdot N_{JI}(\bar{\ell},\ell,S) +\ldots
\label{geo.ser}
\end{align}
 Together with this expansion one also has the standard chiral one (in which a unitarity loop function $g(p^2)$ counts as 
 ${\cal O}(p)$, because of the factor $p$ in Eq.~\eqref{def.g}, despite it is multiplied by the large nucleon mass). 
In this way, for determining $N_{JI}^{(n)}$  one has to match the ${\cal O}(p^n)$ 
ChPT calculation of a $NN$ partial wave with at most $n$ two-nucleon reducible loops with Eq.~\eqref{geo.ser}, 
where $N_{JI}$ is also expanded up to the considered order  
 \begin{align}
 N_{JI}=\sum_{m=0}^n N_{JI}^{(m)}~.
 \label{sum.ord}
 \end{align} 
Here, the chiral order is indicated by the superscript.
 Thus, at lowest order  $N_{JI}^{(0)}(\bar{\ell},\ell,S)$ is given  by 
  the calculation in ChPT at ${\cal O}(p^0)$ without any two-nucleon reducible loop 
(the first diagram in Fig.~\ref{fig:sum} from left to right). 
At ${\cal O}(p)$ the new contribution is the two-nucleon reducible part of the second diagram in the same figure, 
that for a given partial wave is denoted by $L^{(1)}_{JI}(\bar{\ell},\ell,S)$. 
It corresponds to the reducible part of the first iteration of the one-pion exchange plus local vertices.   
Writing $N_{JI}=N_{JI}^{(0)}+N_{JI}^{(1)}+{\cal O}(p^2)$, and matching Eq.~\eqref{geo.ser}  
up to order $g$ with the sum of the first two diagrams in Fig.~\ref{fig:sum} one has
\begin{align}
N_{JI}^{(0)}+N_{JI}^{(1)}-N_{JI}^{(0)}\cdot g \cdot N_{JI}^{(0)}+{\cal O}(p^2)=N_{JI}^{(0)}+L^{(1)}_{JI}+{\cal O}(p^2)~,
\label{Eq.a10}
\end{align}
 with the result
 \begin{align}
N_{JI}^{(1)}=L^{(1)}_{JI}+N_{JI}^{(0)}\cdot g \cdot N_{JI}^{(0)} ~.
\label{Eq.a1} 
\end{align}
Notice that in the expansion of Eq.~\eqref{geo.ser} each factor of the kernel $N_{JI}(\bar{\ell},\ell,S)$ 
multiplies the loop function $g$ with its value on-shell. 
This is why in Eq.~\eqref{Eq.a10} we have $-N_{JI}^{(0)}\cdot g \cdot N_{JI}^{(0)}$ for the first order in $g$. 
This result is then subtracted to the function $L^{(1)}_{JI}$ in Eq.\eqref{Eq.a1} (which guarantees the absence of 
LHC in $N_{JI}$). 
The difference is incorporated in the interaction kernel $N_{JI}(\bar{\ell},\ell,S)$,  
that can be improved order by order. 

At  ${\cal O}(p^2)$ new contributions arise which require the calculation of the irreducible part of the box diagram 
in Fig.~\ref{fig:sum} and the reducible part of the second iteration of the wiggly line, last diagram of Fig.~\ref{fig:sum}. 
In addition there are also chiral counterterms from the NLO quartic nucleon Lagrangian and  two-nucleon irreducible 
pion loops \cite{kaiser,epe,entem}. 
If we denote all these new contributions by $L_{JI}^{(2)}(\bar{\ell},\ell,S)$, 
projected in the corresponding partial wave,  one ends with 
\begin{align}
N_{JI}^{(2)}=L_{JI}^{(2)}+N_{JI}^{(1)}\cdot g\cdot N_{JI}^{(0)}
+N_{JI}^{(0)}\cdot g \cdot N_{JI}^{(1)}-N_{JI}^{(0)}\cdot g\cdot N_{JI}^{(0)}\cdot g \cdot N_{JI}^{(0)}~.
\label{Eq.a2}
\end{align}
That is,  we are  subtracting to $L_{JI}^{(2)}$ the two-nucleon reducible contributions obtained from Eq.~\eqref{geo.ser}  
up to ${\cal O}(p^2)$, in the UChPT expansion of the interaction kernel $N_{JI}(\bar{\ell},\ell,S)$.

 The resulting $N_{JI}$, Eq.~\eqref{sum.ord}, is then substituted in the non-perturbative Eq.~\eqref{master} giving 
the full partial-wave amplitudes $T_{JI}(\bar{\ell},\ell,S)$. 
 One should also stress that Eq.~\eqref{master} is an algebraic one, 
which simplifies tremendously the numerical burden for in-medium calculations.  

An important point is that $N_{JI}$ obeys a non-linear integral equation as derived in
 Ref.~\cite{lacour.180827.3}. Its discontinuity along the LHC is given by
\begin{align}
\hbox{Im}N_{JI}&=\frac{|N_{JI}|^2}{|T_{JI}|^2}\hbox{Im}T_{JI}=|1+g N_{JI}|^2 
\hbox{Im}T_{JI}~,~|\vp|^2<-\frac{m_\pi^2}{4}~.
\label{disc.nji}
\end{align}
 This result can be used to write down a  once-subtracted dispersion
relation for $N_{JI}$,
\begin{align}
N_{JI}(p^2)&=N_{JI}(p_0^2)+\frac{p^2-p_0^2}{\pi}\int_{-\infty}^{-m_\pi^2/4} dk^2\frac{ \hbox{Im}T_{JI}(k^2)\,|1+g(k^2) N_{JI}(k^2)|^2}{(k^2-p^2-i\epsilon)(k^2-p_0^2)}~,
\label{dis.nji}
\end{align}
more subtractions can also be taken.  
 Since  $N_{JI}$ was fixed above by matching with a chiral expansion, the LHC is treated perturbatively. 
Therefore, the most convenient and consistent choice for $g_0$ is such that the function $g(p^2)$ has a zero in the low-energy part 
of the LHC for $p\simeq i\, m_\pi$. 
In this way, the iteration of the LHC contribution from Eq.~\eqref{dis.nji} is expected to have the smallest impact at low 
energies. This implies from Eq.~\eqref{def.g} 
that within our approach the expected value for $g_0$ is  negative and around 
\begin{align}
\label{180830.6}
g_0&\simeq -\frac{mm_\pi}{4\pi}\sim -0.55~ m_\pi^2~.
\end{align}

We now concentrate on fixing the constants $C_S$ and $C_T$ from the local quartic nucleon Lagrangian, Eq.~\eqref{lnn}. 
These constants and $g_0$ are the only free parameters that enter in the evaluation of the $NN$ scattering amplitudes 
from Eq.~\eqref{master} up to ${\cal O}(p)$. 
The counterterms $C_S$ and $C_T$  are fixed by considering  the S-wave $NN$ scattering 
lengths $a_t$ and $a_s$ for the triplet and singlet S-waves, respectively. 
At ${\cal O}(p^0)$ we have 
\begin{align}
\left.T_{JI}(\bar{\ell},\ell,S)\right|_{\text{{\tiny LO}}}=\left[I + N^{(0)}_{JI} \cdot g \right]^{-1}\cdot N^{(0)}_{JI}~,
\label{lo.tnn}\end{align}
where LO stands for leading order. 
Note that the OPE amplitudes, Eq.~\eqref{1pi.gen}, 
vanish at the $NN$ threshold  because they depend quadratically on the nucleon three-momentum. 
For the partial-wave amplitudes\footnote{For general considerations the already introduced notation $T_{JI}(\bar{\ell},\ell,S)$ is employed. Specific partial waves are denoted by the standard spectroscopic notation $T(^{2S+1}\ell_J)$.}
$^1S_0$ and $^3S_1$ at threshold  one has from Eq.~\eqref{lo.tnn},
\begin{align}
T(^1S_0)&=\frac{-(C_S-3C_T)}{1-g_0(C_S-3C_T)}~,\nn\\
T(^3S_1)&=\frac{-(C_S+C_T)}{1-g_0(C_S+C_T)}~.
\label{t.th}
\end{align}
The term $-(C_S-3C_T)$ for $N_{01}$ is a factor 2 smaller than $T_{NN}^{c}(^1S_0)$ in Eq.~\eqref{a.local}, 
 and similarly also for the triplet case, because $N_{JI}$ is given by the direct term.
The resulting expressions for the  scattering lengths from Eq.~\eqref{t.th} are 
\begin{align}
\frac{1}{a_s}&=\frac{2\Lambda}{\pi}+\frac{4\pi/m}{C_S-3C_T}~,\nn\\
\frac{1}{a_t}&=\frac{2\Lambda}{\pi}+\frac{4\pi/m}{C_S+C_T}~.
\label{Eq.1as}
\end{align}
So that
\begin{align}
C_S&=\frac{\pi}{m}\frac{3/a_s+1/a_t-8\Lambda/\pi}{(1/a_s-2\Lambda/\pi)(1/a_t-2\Lambda/\pi)}~,
\nn\\
C_T&=\frac{\pi}{m}\frac{1/a_s-1/a_t}{(1/a_s-2\Lambda/\pi)(1/a_t-2\Lambda/\pi)}~.
\label{cs.ct}
\end{align}
One of the characteristics of $NN$ scattering are the large absolute values of the S-wave scattering lengths 
$a_s=(-23.758\pm 0.04)$~fm and $a_t=(5.424\pm 0.004)$~fm. 
For typical values of $\Lambda$,  $\Lambda\gg |1/a_s|$, $1/a_t$, and then 
$|C_S|\simeq 2\pi^2/m\Lambda \gg |C_T|={\cal O}(\pi^3/a_t/m\Lambda^2)$. 
Because of the introduction of the subtractions constant $g_0$, the low-energy counterterms $C_S$ and $C_T$ do not diverge for 
$a_s$, $a_t\to \infty$. 
In this way, $\Lambda$ is a new scale that adds to the inverse of the scattering lengths so that their sum, 
the one that appears for determining the values of $C_S$ and $C_T$, Eq.~\eqref{cs.ct}, has a natural  size. 
Thus, this procedure  gives rise to values of the chiral counterterms $C_S$ and $C_T$ that are 
consistent with their ascribed ${\cal O}(p^0)$ scaling. 

Indeed, taking into account that: i) Eq.~\eqref{cs.ct} is the same as in the pionless $NN$ EFT, 
because the pion-exchange contribution to $N_{JI}^{(0)}$ vanishes at threshold. 
ii) Naive dimensional analysis implies that $C_{S},\,C_T\sim 4\pi/mQ$, with $Q$ 
the expansion scale for the pionless EFT because of i). 
It follows then that $\Lambda$ should be comparable to $Q$ so that $\Lambda\sim Q ={\cal O}(m_\pi)$. 
Note also that the order of $C_S$  in Eq.~\eqref{cs.ct} is fixed by the product $m\Lambda$, 
which is ${\cal O}(p^0)$  for $\Lambda={\cal O}(p)$ because of the largeness of the nucleon mass. 
This is completely analogous to having counted before the loop function $g(p^2)$ as  ${\cal O}(p^0)$ because 
of the factor $m p$ in Eq.~\eqref{def.g}.
 Let us also recall that, from the consistency requirement of treating perturbatively the LHC within our approach, 
the estimate $\Lambda\sim m_\pi$ is required by our previous estimate of $g_0$ in Eq.~\eqref{180830.6}.
 One has to stress that only local terms and OPE contributions enter in 
 the calculation of   $N^{(0)}_{JI}(\bar{\ell},\ell,S)$. 
 This is certainly too simplistic in order to accurately describe the $NN$ phase shifts, although it might be 
 sensible for capturing the bulk properties in nuclear matter, as explicit applications show \cite{lacour.180827.3,llanes.180827.1}.

\subsection{In-medium $NN$ interactions}

The nucleon propagator in  nuclear matter, $G(k)_{i_3}$, contains both the free and the in-medium contributions \cite{fetter}, 
\begin{align}
G(k)_{i_3}&=\frac{1}{k^0-E(\vk)+i\epsilon}+2\pi i \, \theta(\xi_{i_3}-|\vk|)\delta(k^0-E(\vk))~,
\label{nuc.pro}
\end{align}
In this equation the subscript $i_3$ refers to the third component of isospin of the nucleon, 
so that, $i_3=+1/2$ corresponds to the proton and $-1/2$ to the neutron, 
and the symbol $\xi_{i_3}$ is the Fermi momentum of the Fermi sea for the corresponding nucleon. 
Equivalently, we also use $\xi_p=\xi_{+1/2}$ and $\xi_n=\xi_{-1/2}$ for the proton and neutron Fermi momenta.
We consider that isospin symmetry is conserved so that all the nucleon and pion masses are equal.  
 The Eq.~\eqref{nuc.pro} can also be written equivalently as
\begin{align}
G(k)_{i_3}&=\frac{\theta(|\vk|-\xi_{i_3})}{k^0-E(\vk)+i\epsilon} + \frac{\theta(\xi_{i_3}-|\vk|)}{k^0-E(\vk)-i\epsilon}~.
\label{nuc.prob}
\end{align}
The first (second) term on the right hand side of this equation is known as the particle (hole) part of the nucleon propagator. 
One can gather in a matrix notation the proton and neutron propagators as
\begin{align}
G(k)=&\sum_{i_3}\left(\frac{1}{2}+i_3\tau_3\right)G(k)_{i_3} \nn\\
=&\left( \frac{1+\tau_3}{2} \theta(\xi_p-\mk)+\frac{1-\tau_3}{2}\theta(\xi_n-\mk)\right)\frac{1}{k^0-E(\vk)-i\epsilon}
\nn\\
+&\left( \frac{1+\tau_3}{2} \theta(\mk-\xi_p)+\frac{1-\tau_3}{2}\theta(\mk-\xi_n)\right)\frac{1}{k^0-E(\vk)+i\epsilon}~,
\label{nuc.pro.iso}
\end{align}
or in the equivalent manner
\begin{align}
G(k)&=\frac{1}{k^0-E(\vk)+i\epsilon}+ 2\pi i \, \delta(k^0-E(\vk))
\left( \frac{1+\tau_3}{2} \theta(\xi_p-\mk)+\frac{1-\tau_3}{2}\theta(\xi_n-\mk)\right)\nn~.
\label{nuc.pro.iso.2}
\end{align}

When calculating a loop function in the nuclear medium Ref.~\cite{lacour.180827.3} uses the notation $L_{ij}$, 
where $i$ indicates the number of two-nucleon states in the diagram (0 or 1) and $j$ the number 
of pion exchanges (0, 1 or 2). 
In addition, one also uses  $L_{ij,f}$, $L_{ij,m}$ and  $L_{ij,d}$, 
with the subscripts $m$ and $d$ indicating one or two Fermi-sea insertions from the nucleon propagators
 in the medium, respectively. 
The subscript $f$ refers to the ``free" part and therefore it does not involve any Fermi-sea insertion.  
The labels $f$, $m$ and $d$ originate because the nucleon propagator in the nuclear medium contains both 
a free and an in-medium part, the last proportional to the Dirac delta function in Eq.~\eqref{nuc.pro}. 
In this notation  $g(p^2)=L_{10,f}$ and its in-medium counterpart is $L_{10}$. 
These functions are calculated in the Appendix \ref{app:l10}.

We use the same Eq.~\eqref{master} but now the function $g(p^2)$ is substituted by $L_{10}$. 
The same process as previously discussed is then used to fix $N_{JI}$ in the medium. 
At lowest order they can be easily obtained from our previous result in the vacuum since 
the only modification without  increasing the chiral order is by using the corresponding 
nucleon propagator in the medium, which is directly accomplished by replacing $g(p^2)$ by $L_{10}$. 
Note that  any further in-medium contribution requires to increase $V_\rho$ by at least one unit, 
which then increases, at a minimum, the chiral order one more unit, cf. Eq.~\eqref{fff}. 
The modification of the meson propagators (both for heavy mesons or pions) by the inclusion of an IGV 
 increases the chiral order by two units at least. 
However, the modification of the enhanced nucleon propagators with one IGV 
only increases the order by one more unit and these contributions must be kept when 
calculating the $NN$ scattering amplitudes at NLO. 
The difference stems from the fact that a lowest-order modification of a meson propagator requires  
the inclusion of an IGV with two meson lines which, according to Eq.~\eqref{fff}, increases the power counting 
by two units. 
However, for the baryonic line we would have only one unit of increase because the introduction 
of the extra IGV.

Then, Eq.\eqref{master} for the $NN$ partial-wave in the nuclear medium becomes 
\be
T_{JI}^{I_3}(\bar{\ell},\ell,S)=\left[I + N^{I_3}_{JI}(\bar{\ell},\ell,S) 
\cdot L_{10}^{I_3} \right]^{-1}\cdot N^{I_3}_{JI}(\bar{\ell},\ell,S)~,
\label{l10fa3}
\ee
and $N_{JI}^{I_3}=N_{JI}^{(0)}$ at LO. 
We have included the superscript $I_3$ in Eq.~\eqref{l10fa3}, which corresponds to the third component of the total isospin of the 
two nucleons involved in the scattering process. 
This is due to the fact that in the nuclear medium the Fermi momenta of the neutrons and protons might be different. 
In this way, $L_{10,m}$ and $L_{10,d}$ depend on weather one has two protons, neutrons or a proton and a neutron as intermediate states. 
As a result, a $NN$ partial wave in the nuclear medium depends on the total charge of the intermediate state. 
Of course, this is not the case for the $NN$ interactions in vacuum where they only depend on the total isospin, 
but not on its third component. 
Let us also stress that the absolute value of the total isospin of the $NN$ state, $I$,
 is a good quantum number and does not mix because of the $NN$ interactions. 
The function $L_{10}$  conserves $I$, because it is symmetric under the exchange of the two nucleons, 
though it depends on the charge (or third component of the total isospin) of the intermediate state. 
This is a general rule, all the $I_3=0$ operators are symmetric under the exchange $p\leftrightarrow n$, 
so that they do not mix isospin representations with different exchange symmetry properties.


\section{Nuclear-matter energy density}
\label{sec.180829.1}

\begin{figure}
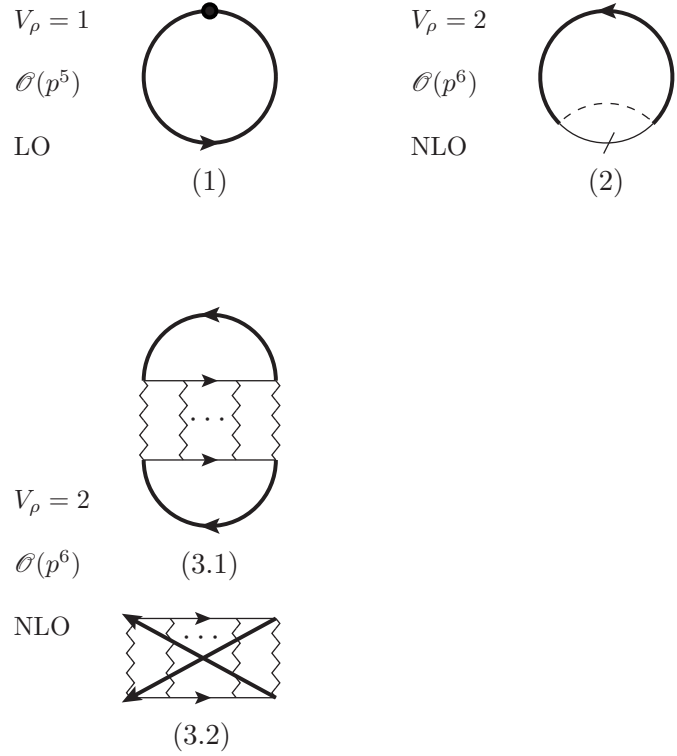

\begin{center}
\begin{axopicture}(200,230)(0,-90)
\Text(-30,172)[l]{{\small \begin{tabular}{l}
        $V_\rho=1$\\
        \\
        ${\cal O}(p^5)$\\
        \\
        LO
        \end{tabular}
  }}
\SetWidth{1.5}
\Arc[arrow,arrowpos=0.75](50,175)(25,0,360) 
\GCirc(50,200){2.5}{.1} 
\Vertex(50,200){1.5} 
\Text(50,135)[c]{(1)} 
\Arc[arrow](200,175)(25,315,225) 
\SetWidth{0.5}
\Arc(200,175)(25,225,315) 
\Line(199,146)(203,154) 
\Arc[dash](200,140)(25,50,130) 
\Text(200,135)[c]{(2)} 
\Text(120,172)[l]{{\small \begin{tabular}{l}
        $V_\rho=2$\\
        \\
        ${\cal O}(p^6)$\\
        \\
        NLO
        \end{tabular}
  }}
\SetWidth{1.5}
\Arc[arrow](50,60)(25,0,180) 
\Arc[arrow,clockwise](50,30)(25,0,180) 
\SetWidth{0.5}
\Line[arrow,arrowpos=0.5](25,30)(75,30)
\Line[arrow,arrowpos=0.5](25,60)(75,60)
\ZigZag(25,30)(25,60){1.5}{4.0}
\ZigZag(75,30)(75,60){1.5}{4.0}
\ZigZag(40,30)(40,60){1.5}{4.0}
\ZigZag(60,30)(60,60){1.5}{4.0}
\Text(50,45)[c]{{\large $\cdots$}}
\Text(50,-10)[c]{(3.1)} 
\Line[arrow,arrowpos=0.5](20,-30)(75,-30) 
\Line[arrow,arrowpos=0.5](20,-60)(75,-60)
\ZigZag(20,-30)(20,-60){1.5}{4.0}
\ZigZag(75,-30)(75,-60){1.5}{4.0}
\ZigZag(35,-30)(35,-60){1.5}{4.0}
\ZigZag(60,-30)(60,-60){1.5}{4.0}
\SetWidth{1.5}
\Line[arrow,arrowpos=1](75,-30)(20,-60)
\Line[arrow,arrowpos=1](75,-60)(20,-30)
\Text(47.5,-37.5)[c]{{\large $\cdots$}}
\Text(47,-75)[c]{(3.2)} 
\Text(-30,-10)[l]{{\small \begin{tabular}{l}
        $V_\rho=2$\\
        \\
        ${\cal O}(p^6)$\\
        \\
        NLO
        \end{tabular}
  }}
\end{axopicture}
\caption{{\small
  The set of diagrams for the evaluation of the energy density in nuclear matter up to and including
two-nucleon interactions in the nuclear medium. The diagram (1) is the kinetic energy, (2) represents the nucleon 
self-energy due to a pion loop [it involves one Fermi-sea insertions only  
so as not double-counting with the diagrams in (3)]. Finally, (3.1) and 
(3.2) are the contributions due to the direct and exchange two-nucleon interactions, in this order,
with at least two in-medium Fermi-sea insertions. The diagrams in (3) are the Hartree (3.1) and Fock (3.2) diagrams.}
\label{fig.180829.1}}
\end{center}
\end{figure}

In this section we consider the energy per nucleon in nuclear matter, ${\cal E}/\rho$, with ${\cal E}$ corresponding to the energy density.
For this case we have to apply Eq.~\eqref{fff} without external legs, and the contributions 
up to NLO are depicted in Fig.~\ref{fig.180829.1}. From left to right and top to 
bottom in this figure, the  three types of contributions  are denoted
  as ${\cal E}_1$, ${\cal E}_2$ and ${\cal E}_3$, in this order, and are evaluated in Ref.~\cite{lacour.180827.3}. 
The LO diagram (1) counts as ${\cal O}(p^5)$ and the NLO ones, (2) and (3),\footnote{As a rule, a set of diagrams indicated 
in a figure as $(i.j)$, with different $j$, are denoted collectively as $(i)$.} as ${\cal O}(p^6)$.

The LO term corresponds to the kinetic energy of the nucleons and it is suppressed because it is a recoil correction 
and has a large nucleon mass in the denominator. Its expression is \cite{lacour.180827.3}
\begin{align}
\label{180829.2}
{\cal E}_1&=\frac{3}{10m}\left(\rho_p\xi_p^2+\rho_n \xi_n^2\right)~,
\end{align}
where we have indicate by $\rho_p$ and $\rho_n$ the proton and neutron densities, respectively. 
 They are related to the Fermi momenta of the nucleons by 
 \begin{align}
\label{180829.3}
\rho_i&=\frac{\xi_i^3}{3\pi^2}~,
 \end{align}
 with $i=1$ referring to proton and $i=2$ to neutron.
 
 For the second contribution ${\cal E}_2$, one has the integration over the Fermi seas  
of the in-medium nucleon self-energy due to a pion loop. 
The contribution for the second diagram in Fig.~\ref{fig.180829.1} with an in-medium part in the baryon propagator underneath the 
pion propagator (dashed line) is not considered in (2) because it is accounted for by the diagram (3.2), 
by the contribution involving only one pion line in the $NN$ interaction. 
In this way Ref.~\cite{lacour.180827.3} derives the following expression for ${\cal E}_2$,
\begin{align}
\label{180829.4}
{\cal E}_2&=-2\int\frac{d^3k}{(2\pi)^3}\left[
\theta(\xi_p-|\vk|)+\theta(\xi_n-|\vk|)\right]\Sigma_f^\pi~,
\end{align} 
where $\Sigma_f^\pi$ is the free part of the pion one-loop self-energy of the nucleon. 
Its expression is a well-known result in Heavy-Baryon ChPT \cite{ulfrev}, which is  also calculated in 
Ref.~\cite{lacour.180827.3}, and given by 
\begin{align}
\label{280830.1}
\Sigma_f^\pi&=\frac{3 g_A^2 b}{32\pi^2 f^2}\left[
-\omega+\sqrt{b}\left(i \log\frac{\omega+i\sqrt{b}}{-\omega+i\sqrt{b}}+\pi
\right)
\right]
-\frac{3g_A^2 m_\pi^3}{32\pi f^2}~,
\end{align}
where $\omega$ is the energy of the nucleon (discounted its vacuum rest mass)
and $b=m_\pi^2-\omega^2-i\epsilon$. In the previous 
expression the value of the self-energy at $\omega=0$ is subtracted since we employ the 
physical nucleon mass $m$. 

The calculation of the diagrams (3.1) and (3.2) of Fig.~\ref{fig.180829.1} requires the 
evaluation of the self-interacting expression
\begin{align}
{\cal
E}_3&=\frac{1}{2}\sum_{\sigma_1,\sigma_2}\sum_{\alpha_1,\alpha_2}\int\frac{d^4
k_1}{(2\pi)^4}\frac{d^4k_2}{(2\pi)^4}e^{ik_1^0\eta}e^{ik_2^0\eta}G_0(k_1)_{
\alpha_1}G_0(k_2)_{\alpha_2}T_{NN}(k_1
\sigma_1\alpha_1,k_2\sigma_2\alpha_2|k_1\sigma_1\alpha_1,k_2
\sigma_2\alpha_2)~.
\label{180830.1}
\end{align}
One has in this equation the symmetry factor 1/2 and the sum
over the spin ($\sigma_i$) and isospin ($\alpha_i$) labels. 
At the end of the calculation one should take the limit $\eta\to 0^+$. 
The convergent factors ($e^{i\eta k_i^0}$) are introduced so that every pure vacuum contribution (without any 
in-medium insertion) is eliminated by construction in the calculation \cite{fetter}. 
This follows because the free part of a baryon propagator, Eq.~\eqref{nuc.prob}, has only a cut 
for negative values of the imaginary part. 
In this way, by closing the energy loop integrations along the upper half-plane, thanks to the convergent factors, 
one avoids the purely vacuum contributions. 

It is convenient to employ CM and relative momentum variables, $a$ and $p$, respectively, defined by
\begin{align}
\label{180830.2}
a&=\frac{1}{2}(k_1+k_2)~,\\
p&=\frac{1}{2}(k_1-k_2)~.\nn
\end{align}
We also introduced the variable $A$ (instead of $a^0$)  as the difference
\begin{align}
\label{180830.3}
A&=2m a^0-\mathbf{a}^2~.
\end{align}
Notice that  $A=\vp^2$ in the CM frame. 

We employ in Eq.~\eqref{180830.1} the in-medium $NN$ scattering amplitudes as  calculated at LO, 
for which  the interaction kernel $N_{JI}^{I_3}(\bar{\ell},\ell,S)$ in Eq.~\eqref{l10fa3} only 
depends on $\vp^2$. The unitarity loop function  $L_{10}^{I_3}$, cf. Appendix \ref{app:l10}, 
depends on $|\mathbf{a}|$ and $A$. 
 As a result the integrand in Eq.~\eqref{180830.1} only depends on $p^0$ through the two explicit nucleon propagators, 
and this integration can be readily done with the resulting expression \cite{lacour.180827.3}
\begin{align}
{\cal
E}_3&=-4i\sum_{\sigma_1,\sigma_2}\sum_{\alpha_1,\alpha_2}\int\frac{d^3a}{
(2\pi)^3}
\frac{d^3
p}{(2\pi)^3}\frac{dA}{2\pi}e^{iA\eta}\,T^{\sigma_1\sigma_2}_{\alpha_1\alpha_2}
(\vp,\vall;A)
\Biggr[ \frac{1}{A-\vp^2+i\ep} \nn\\
&-\frac{\theta(\xi_{\alpha_1}-|\vall+\vp|)
+\theta(\xi_{\alpha_2}-|\vall-\vp|)}{A-\vp^2+i\ep}
-2\pi i \delta(A-\vp^2)
\theta(\xi_{\alpha_1}-|\vall+\vp|)\theta(\xi_{\alpha_2}-|\vall
-\vp|)
\Biggl] ~.
\label{180830.4}
\end{align}

\begin{figure}
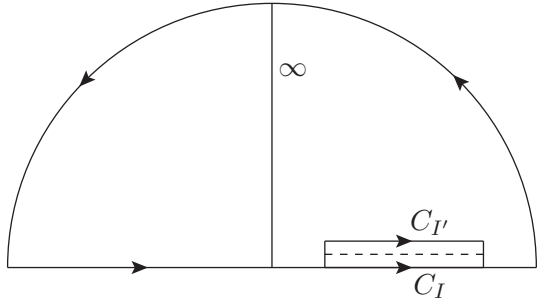

\begin{center}
\begin{tabular}{ll}
\begin{axopicture}(200,100)(0,-20)
\Line[arrow,arrowpos=0.25](0,0)(200,0) 
\Arc[arrow,arrowpos=0.5](100,0)(100,0,90) 
\Arc[arrow,arrowpos=0.5](100,0)(100,90,180) 
\Line(100,0)(100,100) 
\Text(108,75){$\infty$}
\Line[dash](120,5)(180,5) 
\Line[arrow,arrowpos=0.5](120,10)(180,10) 
\Line[arrow,arrowpos=0.5](120,0)(180,0)   
\Line(120,0)(120,10) 
\Line(180,0)(180,10) 
\Text(160,17){$C_{I'}$}
\Text(160,-7){$C_{I}$}
\end{axopicture}
\end{tabular}
\caption{{\small
Contours of integration $C_I$ and $C_{I'}$ on the complex $A$ plane used to
perform the integrals in Eqs.~\eqref{int.a.2} and \eqref{int.a.3}. 
The former contour runs below the
cut (dashed line) and the latter above it. The limits of the cut in $A$ due to
the hole-hole part of $L_{10}$, Eq.~\eqref{l10.b.pp}, are $A(\alpha)$ and $B(\alpha)$.} 
\label{fig:ci}}
\end{center}
\end{figure}


We proceed next to perform the integration in the variable $A$ by closing the integration contour 
with a half circle in the upper half plane of the complex $A$ plane by taking advantage of the 
convergent factor $e^{iA\eta}$.\footnote{Indeed, one does not really need to invoke the presence of this convergent factor because of the  dependence on $A$ of $T^{\sigma_1\sigma_2}_{\alpha_1\alpha_2}(\vp,\vall;A)$.} As discussed above, the cuts due to the particle-particle contribution 
in the unitarity loop function give rise to a cut running parallel to the real axis but with slightly 
negative imaginary part. Therefore, only the hole-hole contribution in $L_{10}^{I_3}$ gives rise to a cut 
with slightly positive imaginary part but of finite extent (for more details see Eq.~\eqref{l10.b.pp} and the discussions that follows there). 
The latter is represented in Fig.~\ref{fig:ci} by the dashed line, 
it extends from $A(\alpha)$ to $B(\alpha)$ with $\alpha=|\mathbf{a}|$, and it is given in   Eq.~\eqref{l10d.exp}.  
The integration contours  $C_I$ and $C_{I'}$ avoid this cut by running slightly 
above and below it. In order to perform the integration in $A$ we make use of the equalities
 \begin{align}
\label{int.a.2}
&\int_{-\infty}^{+\infty}\frac{dA}{2\pi}\frac{e^{iA\eta}}{A-\vp^2+i\epsilon}T^{
\sigma_1\sigma_2}_{\alpha_1\alpha_2}(\vp,\vall;A)=\oint_{C_I}\frac{dA}{
2\pi}\frac{e^{iA\eta}}{
A-\vp^2+i\epsilon}T^{\sigma_1\sigma_2}_{\alpha_1\alpha_2}(\vp,\vall;A)~,\\
&\oint_{C_{I'}}\frac{dA}{2\pi}\frac{e^{iA\eta}}{A-\vp^2+i\epsilon}T^{
\sigma_1\sigma_2}_{\alpha_1\alpha_2}(\vp,\vall;A)=0~.
\label{int.a.3}
 \end{align}
Subtracting Eq.~\eqref{int.a.3} to Eq.~\eqref{int.a.2}  we then have for the integration sought
\begin{align}
\oint_{C_I}\frac{dA}{2\pi}\frac{e^{iA\eta}}{A-\vp^2+i\epsilon}T^{
\sigma_1\sigma_2}_{\alpha_1\alpha_2}(\vp,\vall;A)&-\oint_{C_{I'}}\frac{dA
}{2\pi}\frac{e^{iA\eta}}{
A-\vp^2+i\epsilon}T^{\sigma_1\sigma_2}_{\alpha_1\alpha_2}(\vp,\vall
;A)\nn\\
&=\int_{A(\alpha)}^{B(\alpha)}\frac{dA}{2\pi}\frac{T^{\sigma_1\sigma_2}_{
\alpha_1\alpha_2}(\vp,\vall;A)-T^{\sigma_1\sigma_2}_{\alpha_1\alpha_2}
(\vp,\vall;A+2i\ep)}{A-\vp^2+i\ep
}~.
\label{int.a.4}
\end{align}
The difference of the $T$ matrices is not zero due to the contribution 
to the unitarity cut from the hole-hole part of $L_{10}$. From the partial-wave 
expansion of the $T$ matrix and the calculation of $L_{10}^{I_3}$ in Appendix \ref{app:l10} it results  
\cite{lacour.180827.3}
\begin{align}
&{\cal E}_3=-4\sum_{I,J,\ell,S}\sum_{I_3=-1}^1(2J+1)\chi(S\ell
I)^2\int\frac{d^3a}{(2\pi)^3}
\frac{d^3
q}{(2\pi)^3}\theta(\xi_{\alpha_1}-|\vall+\vq|)\theta(\xi_{\alpha_2}-|\vall-\vq|)\Biggl(
T_{JI}^{I_3}(\vq^2,\vall^2,\vq^2)\nn\\
&+m\int\frac{d^3p}{(2\pi)^3}\frac{1-\theta(\xi_{\alpha_1}-|\vall+\vp|)
-\theta(\xi_{\alpha_2}-|\vall-\vp|)}{\vp^2-\vq^2-i\ep}
\nn\\
&\times
\left[{N_{JI}^{I_3}(\vp^2)}^{-1}+L_{10}^{I_3}(\vall^2,\vq^2)\right]^{-1}
\cdot\left[{N_{JI}^{I_3}(\vp^2)}^{-1}+L_{10}^{I_3}(\vall^2,\vq^2+2i\ep)\right]^{
-1}
\Biggr)_{(\ell,\ell,S)},
\label{e3.trans}
\end{align}
 The isospin index $I_3=\alpha_1+\alpha_2$, and for $I_3=0$ one should take just 
one of the two possible cases with $\alpha_1=-\alpha_2$, $|\alpha_1|=1/2$ for evaluating $L_{10}^{I_3}$. 
The symbol $\chi(S\ell I)$ is defined as
\be
\chi(S\ell I)=\frac{1-(-1)^{\ell+S+I}}{\sqrt{2}}=\left\{
\begin{array}{ll}
\sqrt{2} & \ell+S+I=\hbox{ odd}\\
0 & \ell+S+I=\hbox{ even}
\end{array}~.
\right.
\label{chi}
\ee
An important technical point demonstrated in Ref.~\cite{lacour.180827.3} is that 
${\cal E}_3$ from Eq.~\eqref{e3.trans} is purely real, despite it involves complex 
functions. We do not repeat here the general demonstration given in this reference, which is based on Pauli blocking, 
unitarity and the fact that 
$N_{JI}$ has no RHC (the latter stems only from $L_{10}$). 

\begin{figure}[ht]
\psfrag{E}{{\small  ${\cal E}/\rho$ (MeV)}}
\psfrag{F}{{\small \hspace{0.225cm} ${\cal E}/\rho$ (MeV)}}
\psfrag{N}{{\small \hspace{-1cm} Neutron Matter}}
\psfrag{M}{{\small \hspace{-1.25cm} Symmetric Nuclear Matter}}
\psfrag{rho}{$\begin{array}{c}
\\{\small \hspace{-1cm}\rho~(\hbox{fm}^{-3})}
\end{array}$}
\psfrag{g}{\small $\begin{array}{c}
    g_0\simeq -1.0~m_\pi^2\\
    \widetilde{g}_0=-0.5~m_\pi^2
  \end{array}$}
\psfrag{gg}{\small $g_0\simeq \widetilde{g}_0\simeq -0.65~m_\pi^2$}
\begin{center}
\begin{tabular}{ll}
\epsfig{file=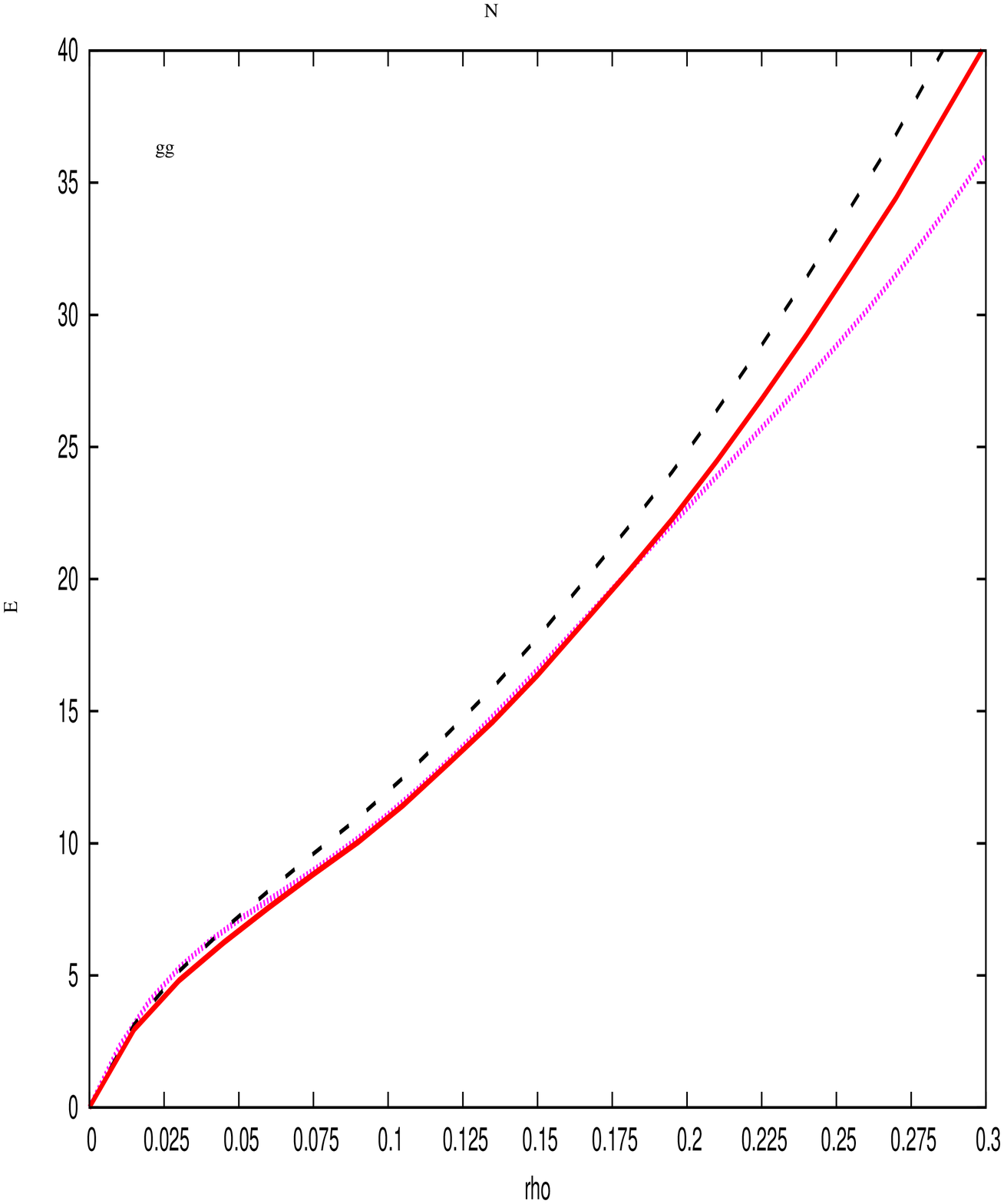,width=.4\textwidth,height=.38\textheight,angle=-0}
&
\epsfig{file=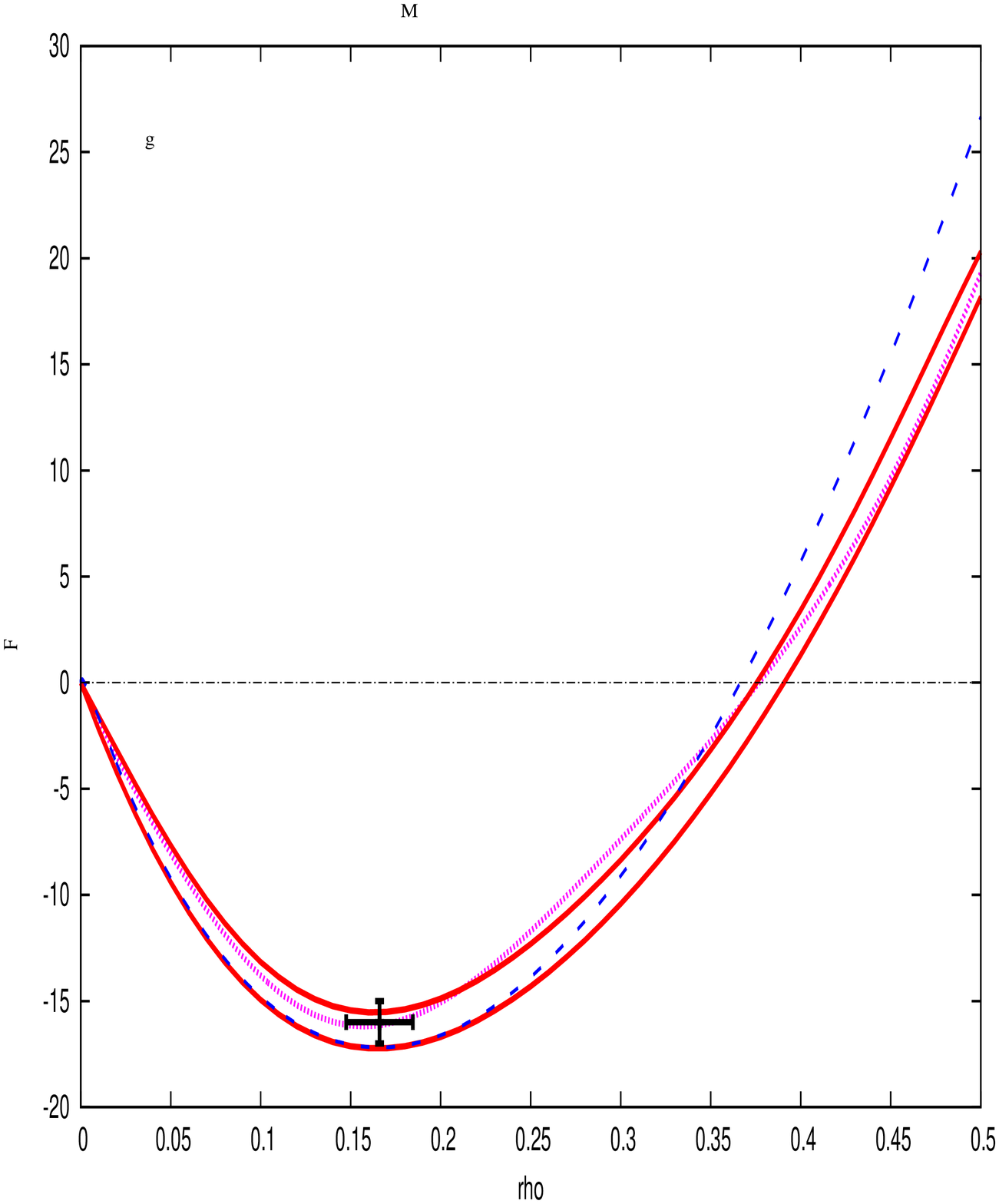,width=.4\textwidth,height=.38\textheight,angle=-0}
\end{tabular}
\end{center}
\vspace{0.2cm}
\caption[pilf]{\protect \small (Color online.) The energy per nucleon ${\cal E}/\rho$ for pure neutron matter (left panel) and 
symmetric nuclear matter (right panel). 
 The (magenta) dotted lines correspond to the results of Ref.~\cite{urbana}. 
Left panel: The (black) dashed line is obtained with $g_0=\widetilde{g}_0=-0.62~m_\pi^2$. 
The (red) solid line employes  slightly different values, $g_0=-0.62~m_\pi^2$ and  $\widetilde{g}_0=-0.65~m_\pi^2$. 
Right panel: The two (red) solid lines correspond from top to bottom to 
 $(g_0,\widetilde{g}_0)=(-0.977,-0.512)~m_\pi^2$ and $(-0.967,-0.525)~m_\pi^2$, in this order.  
The (blue) dashed line is obtained from Eq.~\eqref{param.kw} by adjusting the position of the minimum  
and  its value. 
\label{fig.180830.1}}
\end{figure}

The LO UChPT calculation of $T_{JI}$ implies that the product 
\begin{align}
\Sigma_{p\ell}&=\sum_{\ell'}T_{JI}^{I_3}(\ell,\ell',S;\vp^2,\vall^2,\vq^2)T_{JI}^{I_3}(\ell',\ell,S;\vp^2,
\vall^2,\vq^2)^*~,
\label{180830.5}
\end{align}
which appears in the last integral of Eq.~\eqref{e3.trans}, tends to constant for $\vp^2\to \infty$. 
As a result this integral is linearly divergent. The regularization undertaken in Ref.~\cite{lacour.180827.3} 
consists of adding and subtracting the limiting value for $p^2\to \infty$ 
of the previous sum, denoted by $\Sigma_{\infty \ell}$. The integration of the difference 
$\Sigma_{p\ell}-\Sigma_{\infty\ell}$   is finite because it vanishes as $p^{-2}$ for $p^2\to \infty$. 
The remaining divergence is multiplied by the constant $\Sigma_{\infty \ell}$ and is of the same type 
as the one stemming from the unitarization of the $NN$ amplitudes, cf. Eq.~\eqref{int.g}, which is 
then regularized by including a constant $\widetilde{g}_0$. 
The resulting expression is \cite{lacour.180827.3}
\begin{align}
{\cal E}_3&=4\sum_{I,J,\ell,S}\sum_{\alpha_1,\alpha_2}(2J+1)\chi(S\ell
I)^2 \int\frac{d^3a}{(2\pi)^3}  \frac{d^3 q}{(2\pi)^3} \theta(\xi_{\alpha_1}-|\vall+\vq|)\theta(\xi_{\alpha_2}-|\mathbf{a}-\vq|) 
\Bigl[
-T_{JI}^{I_3}(\ell,\ell,S;\vq^2,\vall^2,\vq^2)\nn\\
& + \widetilde{g}_0 \Sigma_{\infty \ell} -
m \int \frac{d^3p}{(2\pi)^3} 
\Bigl\{
\frac{1-\theta(\xi_{\alpha_1}-|\vall+\vp|) - \theta(\xi_{\alpha_2}-|\vall-\vp|)}{\vp^2-\vq^2-i\ep}
\Sigma_{p\ell}-\frac{1}{\vp^2}\Sigma_{\infty\ell}\Bigr\}
\Bigr]~.
\label{e3.reg.2}
\end{align}
Therefore, we have two constants  $\widetilde{g}_0$ and $g_0$. The latter comes from the 
$NN$ scattering amplitudes, with a value estimated in Eq.~\eqref{180830.6},
 and the former has appeared specifically in the calculation of the in-medium density energy. 
 In principle both parameters should be the same. 
 As a first exercise, let us vary $g_0=\widetilde{g}_0$ in order to improve the description of ${\cal E}/\rho$ (that is, the energy 
 per nucleon) for the
 case of neutron matter, so that our results agree better with the magenta dotted line in the left panel of 
 Fig.~\ref{fig.180830.1} that corresponds to
 the sophisticated many-body calculation of Ref.~\cite{urbana}. The latter employs  realistic $NN$ interactions,
and also includes a free parameter to mimic the three-nucleon ($3N$) interactions. 
 The black dashed line in Fig.~\ref{fig.180830.1} is obtained by employing
 $g_0=\widetilde{g}_0=-0.62~m_\pi^2$, which closely reproduces the results of Ref.~\cite{urbana}, even up to
 rather high densities.  Note as well that this result is obtained with
 a value of $g_0$ which is still very close to its natural size of around $-0.55~m_\pi^2$, cf. Eq.~\eqref{180830.6}. 
 
 However, if we employ the  same value for $g_0=\widetilde{g}_0$ to evaluate ${\cal E}/\rho$  in the case of symmetric nuclear matter 
 the resulting curve has the minimum at its right position, $\rho\simeq 0.16~$fm$^3$, but the value of the energy per baryon is around $-42$~MeV, which is an over-binding by a factor 2.5. 
 Therefore, in order to improve this fact we allow different values for $g_0$ and 
 $\widetilde{g}_0$ so as to mimic higher order contributions (we already discussed at the end of Sec.~\ref{sec:pw} that, because of the size of $\Lambda_\xi$, we expect corrections of  around 30\% from higher order contributions at the nuclear matter saturation density). 
 Once $g_0$ is fitted to 
provide a good reproduction of the properties of the minimum energy per baryon and saturation density in symmetric nuclear 
matter we then obtain a good reproduction of the results of Ref.~\cite{urbana} as well. 
We show  by the two red solid lines in the right panel of Fig.~\ref{fig.180830.1} the results obtained by Ref.~\cite{lacour.180827.3} 
with the values $(g_0 , \widetilde{g}_0)   = (-0.977,-0.512)~m_\pi^2$ and  $(-0.967,-0.525)~m_\pi^2$, 
in this order from top to bottom in the figure.  
These curves run very close to the sophisticated and standard many-body calculation 
of Ref.~\cite{urbana}. Again the value for the subtraction constant $\widetilde{g}_0$ is very close to the expectation from 
Eq.~\eqref{180830.6}, while that for $g_0$ is off by  a factor less than 2 [which is still a number of ${\cal O}(1)$].
At the saturation point one has the minimum value for  ${\cal E}/\rho$ 
with the results $-15.4$ and $-17.1$~MeV, respectively. The experimental
value given by the cross corresponds to $-16\pm 1~$MeV. The densities for the minimum energy per baryon are 
$\rho=0.169$ and $\rho=0.168$~fm$^{-3}$, respectively, compared
to the empirical value $\rho=0.166\pm 0.019$~fm$^{-3}$. 
It is clear  that our results reproduce  the saturation point and agree 
with the calculation of ref.~\cite{urbana} remarkably well. 

 Let us also clarify that the different curves  in the two panels of Fig.~\ref{fig.180830.1} for the results of Ref.~\cite{lacour.180827.3} are not intended to show the intrinsic uncertainty in the calculation. In the case of symmetric nuclear matter the two curves stem 
from the reproduction of the minimum value of ${\cal E}/\rho$ and the saturation density approximately within their error ranges. 
For the neutron matter case the two curves correspond to either keeping the constraint $g_0=\bar{g}_0$ 
or allow for different values (the red solid curve has ($g_0,\widetilde{g}_0)=(-0.62,-0.65)~m_\pi^2$). 
A close reproduction of the results of \cite{urbana} for neutron matter requires that indeed these two scenarios 
 are basically coincident.

The blue dashed line in Fig.~\ref{fig.180830.1} originates by the  parameterization of  
the energy per baryon in symmetric nuclear matter as
\begin{align}
\frac{{\cal E}}{\rho}&=\frac{3\xi^2}{10 m}-\alpha\frac{\xi^3}{m^2}+\beta\frac{\xi^4}{m^3}~.
\label{param.kw}
\end{align}
Interestingly, the parameters $\alpha$ and $\beta$ in the previous equation 
are fixed by adjusting the empirical position and minimum value of the nuclear matter saturation point. 
 The dashed line runs very close to the solid lines determined in Ref.~\cite{lacour.180827.3} 
 and starts to deviate for densities  above $\rho\simeq 0.25$~fm$^{-3}$. 
The nuclear matter incompressibility is given by the double derivative
\begin{align}
K=\left.\xi^2\frac{\partial^2} {\partial \xi^2}\frac{\cal E}{\rho}\right|_{\xi_0}~,
\label{k.exp}
\end{align}
with $\xi_0$ the Fermi momentum at the saturation point. 
From the parameterization in Eq.~\eqref{param.kw} one has $K=259~$MeV, with $\alpha$ and $\beta$ 
determined by using the values of the point with errorbars in the right panel of 
Fig.~\ref{fig.180830.1}. This value is well inside the experimental determination
 $K=250\pm 25~$MeV \cite{globo1}.  
 The resulting nuclear matter incompressibility calculated from Eq.~\eqref{e3.reg.2}  is 
$K=254$ and $233~$MeV, for the upper and lower solid curves, respectively. 
These values are compatible with  the experimental measurement.

\section{Application of the EOS to neutron stars}
\label{sec.180829.1b}

As an application of the energy density ${\cal E}(\rho)$  calculated in Sec.~\ref{sec.180829.1}, we consider the solution 
of the Tolman-Oppenheimer-Volkoff equation \cite{tolman.180902.1,oppen.180902.1} 
for the hydrostatic equilibrium of a cold non-rotating neutron star, as developed in Ref.~\cite{llanes.180827.1}. 
We assume that the neutron star is composed purely of neutrons that interact between them by the 
strong and gravitational interactions. The latter is attractive and
 tends to bound the system to the smallest possible size, being eventually counterbalanced by the 
inter-neutron repulsion at short distances. This compensating mechanism is effective until a maximum 
mass of the star, above which the gravitation attraction could not be overcome by the strong interactions and 
the system would collapse into a black hole. As it is known since the work of Oppenheimer and 
Volkoff \cite{oppen.180902.1}, a  neutron gas with only gravitational interactions  
can support, because of the Pauli exclusion principle,
 a neutron star up to a maximum mass of around 0.7 solar masses ($M_{\odot}$). 
This is the simple, but fundamental, model of a degenerate cold Fermi gas of neutrons. 
In recent years, a neutron star with a mass of 1.97(4) $M_{\odot}$ 
was detected \cite{demorest.180827.1}. To fill the gap from 0.7~$M_{\odot}$ up to 2~$M_{\odot}$ 
an extra repulsion is needed beyond the Pauli exclusion principle. 
The latter implies that making the system more compact requires more energy, since one should 
end up having a bigger density and therefore a bigger Fermi momentum.
Of course, the nuclear interactions 
provide extra and stronger repulsion that allows the increase of the maximum neutron-star 
mass up to around 2~$M_{\odot}$. 

 There are intriguing coincidences that make that this maximum neutron-star mass is only around a factor of 
3 larger than the limit for a degenerate Fermi sea. This is triggered by the fact that the pion coupling 
between two nucleons in the realm of momenta probed in the neutron star (in turn given by the Newton gravitational  
constant $G$, $\hbar$, $c$ and the neutron mass $m$, as it is clear from the cold degenerate Fermi gas)
 is governed by the dimensionless combination of parameters 
 $g_A^2 m_\pi^2/f_\pi^2\simeq 3.5$, with $f_\pi$ the pion weak decay constant in vacuum.  
Another important scale is the mass of the neutron that suppresses the kinetic energy contribution to 
${\cal E}$ and enhances the nuclear one, by giving rise to the relatively low
 scale $\Lambda_{\rm low}\sim \pi (4f_\pi)^2/g_A^2 m\simeq 2 m_\pi$, despite that 
 it does not involve any parameter vanishing in the  chiral limit. 
This type of enhancement is typical of scattering between particles of heavy-mass compared to the typical 
three-momenta in the process.
 
 The two basic equations that were used in the previous study of Ref.~\cite{llanes.180827.1} for the
application of the energy density derived in Ref.~\cite{lacour.180827.3}
 to neutron stars in hydrostatic equilibrium are:
 
 i) The thermodynamical expression giving the pressure $P$ as a function of the density $\rho$, 
\begin{align}
\label{180902.1}
P&=\rho\frac{\partial {\cal E}(\rho)}{\partial \rho}-{\cal E}(\rho)~.
\end{align}
This equation can be deduced by relating the energy density with the 
energy per unit mass $u$, ${\cal E}=u \rho m$, and then 
employing the equation $du=P d\rho/(m \rho^2)$, where we have taken into account that 
the mass density is $m \rho$.

ii) The Tolman-Oppenheimer-Volkoff equation
\begin{align}
\label{180902.2}
\frac{dP(r)}{dr}&=-\frac{G}{r^2}\frac{[{\cal E}(r)+P(r)][M(r)+4\pi r^3 P(r)]}{1-2G M(r)/r}~,
\end{align}
where $r$ is the radial distance and $M(r)$ is the mass contained in a sphere of radius $r$. 
The knowledge of $P$ from Eq.~\eqref{180902.1}, relating this magnitude with $\rho$, 
allows  to solve the previous differential equation by giving an initial value to the pressure 
 at $r=0$ and integrating up to the distance $r=R$ where it vanishes, being $R$ the radius of the star.
 
 \begin{figure}[t]
\begin{center}
\begin{tabular}{ll}
\psfrag{p}{{\small $P~(m_\pi^4)$}}
\psfrag{e}{{\small  $m\rho+{\cal E}~(m_\pi^4)$}}
  \epsfig{file=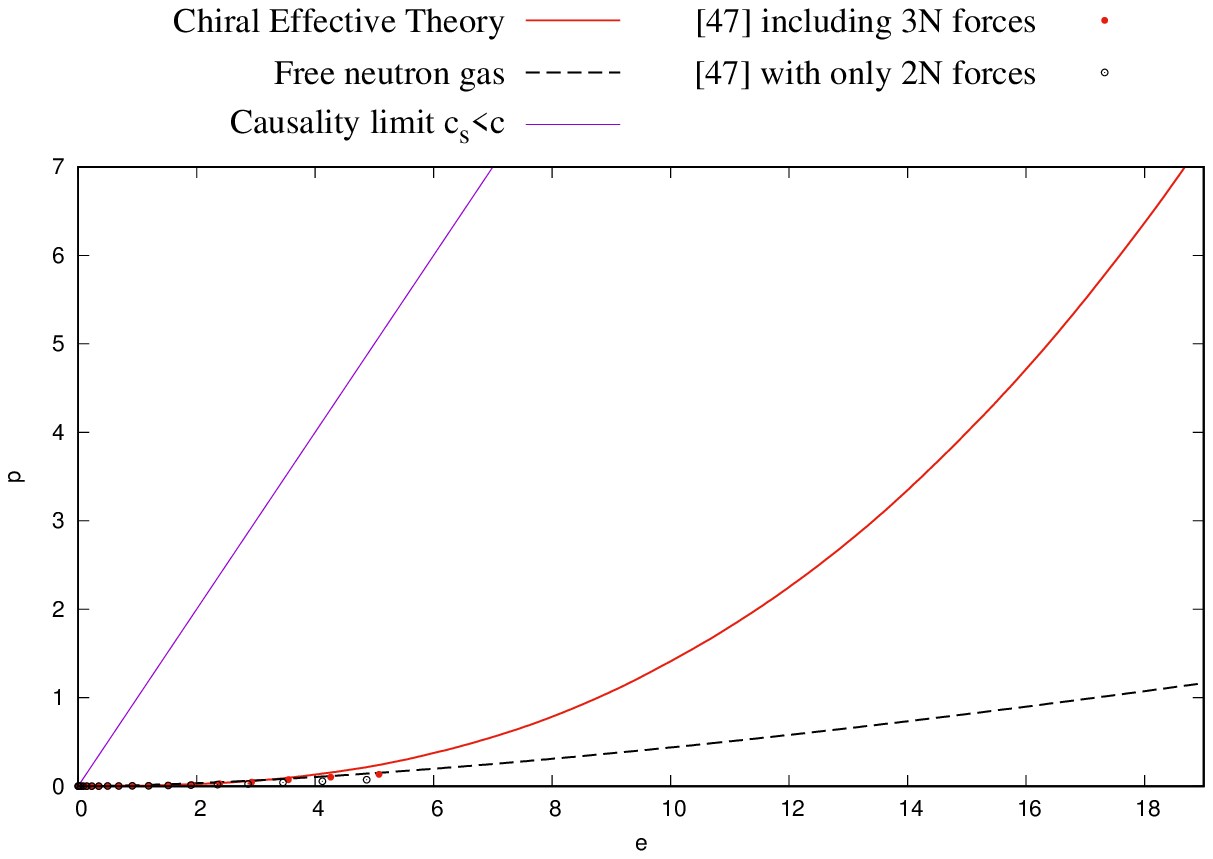,width=.42\textwidth,height=.36\textheight,angle=0} &
\psfrag{r}{{\small  $R$~(km)}}
\psfrag{m}{{\small $M/M_\odot$}}
\epsfig{file=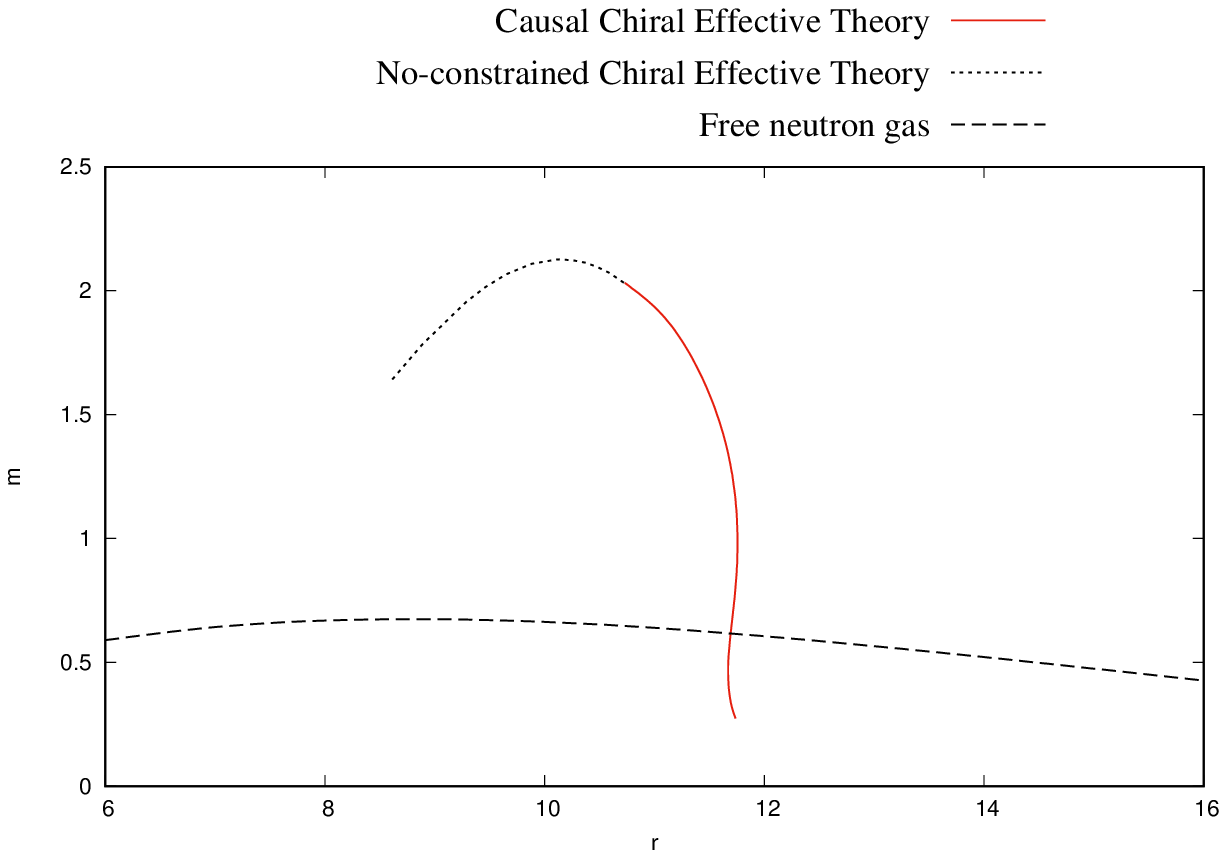,width=.42\textwidth,height=.36\textheight,angle=0}
\end{tabular}
\end{center}
\vspace{0.2cm}
\caption[pilf]{\protect \small 
 Left panel: The pressure $P$ as a function of the total energy density $m\rho+{\cal E}$ 
 in neutron matter. 
The red solid line is the prediction from NLO in-medium ChPT worked out in Ref.~\cite{llanes.180827.1}. 
The black dashed line is the EOS for degenerate cold Fermi gas. The dots come from Ref.~\cite{Hebeler}. 
In addition to the $2N$ forces, the red filled circles include $3N$ interactions 
while the empty circles do not. 
The region to the left of the violet solid line is forbidden because it violates the causality limit,
 $c_s^2=\partial P/\partial m \rho<c^2$. 
This line is given by $P(m\rho)=m\rho$ (in our units $c=1$). Notice also that  the causality constraint
applies to the slope of the pressure in any curve of the figure (since $m\rho\gg {\cal E}$),
independently of whether they lie to the right of the violet solid line.
 The right panel is a mass-radius plot.  
The red solid and black dotted lines are the results obtained by making use of the EOS from the NLO in-medium ChPT 
\cite{llanes.180827.1}. The curve corresponding to the EOS of a degenerate cold Fermi gas is the black dashed line. 
For more details see the text. 
\label{fig.180902.1}}
\end{figure} 

We show in the left panel of Fig.~\ref{fig.180902.1} the pressure as a function of the energy density. 
The EOS that follows from the calculation within the NLO in-medium ChPT of ${\cal E}(\rho)$ \cite{lacour.180827.3},  
and worked out in Ref.~\cite{llanes.180827.1}, is indicated 
by the red solid line, while the black dashed  line corresponds to the cold degenerate Fermi-sea limit. 
 In the figure the left side of the violet line is a  region excluded because the sound velocity 
 ($c_s^2=\partial P/\partial m \rho$) would be larger than the speed of light. Apart from this, since this
 constraint really refers to the slope of $P(m\rho)$ (notice that $m\rho\gg {\cal E}$), it also excludes the equations of state 
with a pressure raising too fast as a function of the energy density for large Fermi momentum.
In addition, we have also shown two sets of dots that correspond to the results of Ref.~\cite{Hebeler}. 
 Both of them include $2N$ interactions but, in addition,  the red filled circles take into account $3N$ interactions 
 while the black empty circles do not 
  (more details are given in the many-body calculation of Ref.~\cite{hebeler.181230.1}). 
The $3N$ interactions make stiffer the resulting EOS. 

 Next, Ref.~\cite{llanes.180827.1} obtained the mass-radius relation given by the red 
solid curve  in the right panel of Fig.~\ref{fig.180902.1} from  
the solution of the Tolman-Oppenheimer-Volkoff equation with the EOS provided 
at NLO by in-medium ChPT \cite{lacour.180827.3}.  
It clearly allows for a neutron-star mass slightly above  two solar masses, with a maximum of 2.15~$M_{\odot}$. 
The black dotted line is obtained by entering in a region that involves relatively high Fermi momenta, larger than 600~MeV, 
above which the  chiral effective field theory is expected to break. 
Indeed, the NLO in-medium calculation of ${\cal E}(\rho)$ 
of Ref.~\cite{lacour.180827.3} starts giving rise 
to a sound velocity larger than the speed of light for such large values of $\xi_n$. 
The allowed neutron-star mass then reaches the highest maximum value of 
2.25 $M_\odot$. 

We can also compare our prediction for the mass-radius relation in Fig.~\ref{fig.180902.1} with 
the most recent determinations from the event GW170817 observed by the LIGO and Virgo Collaborations 
\cite{ligo.180827.1}. 
 Reference~\cite{ruiz.180902.1} has derived upper bounds  for the maximum mass of a cold and spherical 
  neutron star, $M_{\rm max}$, 
  from the resulting kilonova observations  associated to  GW170817:
\begin{align}
 \label{180902.3}
M_{\rm max}&=(2.16-2.28)M_\odot\pm 0.23~M_\odot,
\end{align}
The results obtained in Ref.~\cite{llanes.180827.1} corresponding to the solid and dotted 
lines of Fig.~\ref{fig.180902.1}, are in agreement with these upper bounds. 
The solid line, as explained above, does always involve Fermi momenta small enough so that 
$c_s<1$ in the whole process of solving the Tolman-Oppenheimer-Volkoff equation 
and it gives $M_{\rm max}=2.15~M_{\odot}$.  Even the maximum value on the dotted line 
implies $M_{\rm max}=2.25~M_{\odot}$, which still is compatible with the values in 
Eq.~\eqref{180902.3}. 

Another constraint is put forward by Ref.~\cite{annala.180827.1}, which concludes that 
the radius of a neutron star of mass $1.4\,M_\odot$, $R(1.4 \, M_{\odot})$,
 should be between 9.9 and 13.6~km. Reference~\cite{annala.180827.1} obtains this range 
by taking into account: i) The existence of 
two solar-mass neutron stars. ii) The 
tidal deformability of a neutron star mass of $1.4~M_{\odot}$, $\Lambda(1.4 M_{\odot})$,
 should be smaller than 800, as established by the LIGO and Virgo Collaborations in Ref.~\cite{ligo.180827.1} 
 from the GW170817 event. With this information, Ref.~\cite{annala.180827.1} constrains 
a rather generic family of equations of 
 state that are matched at low and high densities with those derived by taking ingredients 
 from chiral effective field theory  \cite{sch.180902.1}
and the one from a N$^2$LO perturbative QCD calculation for cold quark matter \cite{pqcd.180902.1}, in this order.  
 From the red solid line  in the right panel of Fig.~\ref{fig.180902.1}  we have that $R(1.4\,M_{\odot})=12.3$~km, well inside 
the previous interval from Ref.~\cite{annala.180827.1}. Even if allowing for $\Lambda(1.4\,M_{\odot})<400$, as 
roughly suggested by the 50\% of the contours in Fig.~5 of Ref.~\cite{ligo.180827.1},  
the constraint $R(1.4\,M_{\odot})<12.5$~km can then be concluded \cite{annala.180827.1}, 
 which is  also fulfilled by our curve.

A main point of Ref.~\cite{llanes.180827.1} is to use the knowledge of a sound EOS to put a constraint 
on the value of the Newton gravitational constant in an environment of strong gravitational field. The 
idea is that if $G$ increases   for a given EOS, the system would become more compact due 
to the stronger gravitational attraction and, at some point, the maximum allowed mass for a neutron star 
would be smaller than the actual mass of 1.97(4) $M_{\odot}$ 
	detected in Ref.~\cite{demorest.180827.1}. 
It was found in Ref.~\cite{llanes.180827.1} that,
by employing the NLO in-medium ChPT calculation of ${\cal E}(\rho)$ from Ref.~\cite{lacour.180827.3},  
this argumentation requires that 
$G$ cannot be more than 12\% larger compared to its value on the surface of the Earth
at the 95\% confidence  level in a 2~$M_\odot$ neutron star. 
As shown in the Fig.~5 of the same reference, the predicted gravitational acceleration in the star 
 reaches a typical magnitude of $10^{13}~m/s^2$, therefore twelve orders of magnitude larger than the one 
on the surface of the Earth.

\section[The in-medium temporal pion-decay constant, the quark condensate and the pion self energy]{The in-medium temporal pion-decay constant, the quark condensate and the pion self energy}
\label{sec.180903.1}

Two other important quantities to characterize the nuclear-matter state 
 are the pion-decay constant, $f_t$, associated with the temporal component of the axial-vector current,  
 and the quark condensate. 
These quantities were studied in Refs.~\cite{lacour.180827.3} and \cite{lacour.180827.4}   
 by applying in-medium ChPT up to NLO order. 
Before these works, Ref.~\cite{annp} also considered 
the evaluation of these quantities, though non-perturbative $NN$ interactions were not included. 
However, as explained in Refs.~\cite{lacour.180827.3}, the latter contributions exactly cancel 
for the temporal-component coupling $f_t$ and the pion self-energy, so that  
the expressions obtained in Ref.~\cite{annp} fully hold up to NLO. In the case of the 
in-medium quark condensate  strong cancellations occur, though some NLO corrections to the result of 
Ref.~\cite{annp} remain from the $NN$ interactions, as worked out in Ref.~\cite{lacour.180827.4}.

\begin{figure}
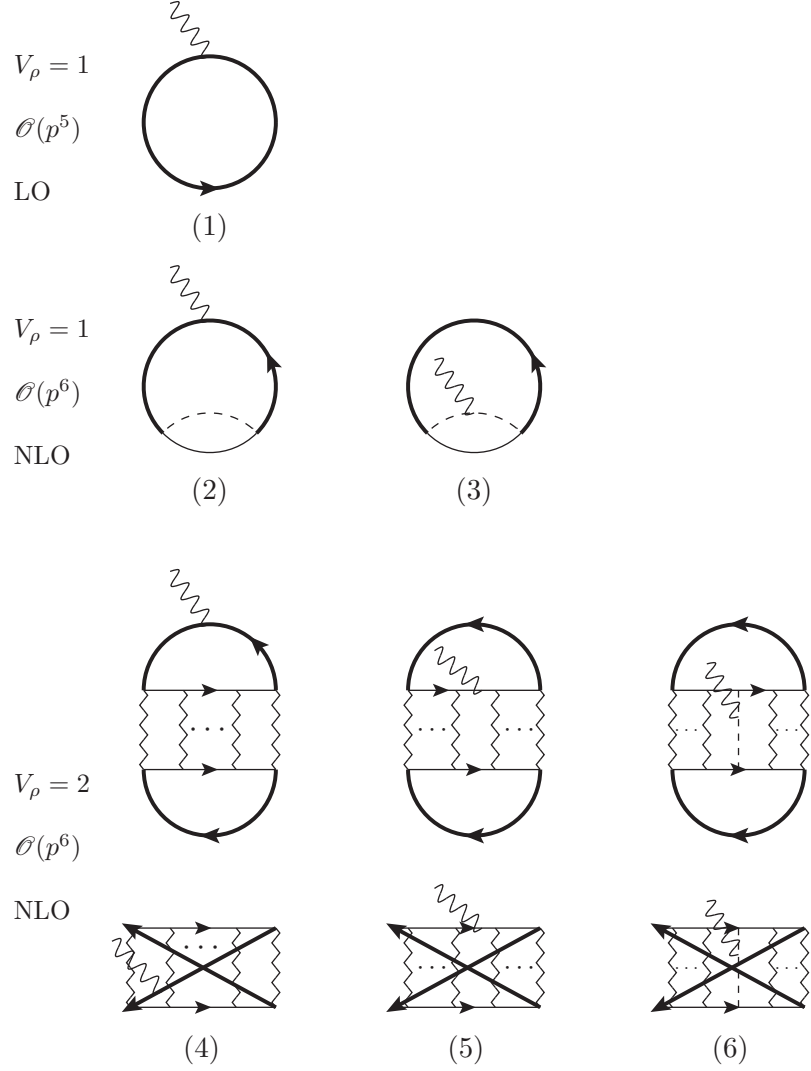

\begin{center}
\begin{axopicture}(200,330)(0,-90)
\Text(-30,272)[l]{{\small \begin{tabular}{l}
        $V_\rho=1$\\
        \\
        ${\cal O}(p^5)$\\
        \\
        LO
        \end{tabular}
  }}
\SetWidth{1.5}
\Arc[arrow,arrowpos=0.75](50,275)(25,0,360) 
\SetWidth{0.5}
\Photon(50,300)(35,320){3}{4}
\Text(50,235)[c]{(1)} 
\SetWidth{1.5}
\Arc[arrow,arrowpos=0.25](50,175)(25,315,225) 
\SetWidth{0.5}
\Arc(50,175)(25,225,315) 
\Arc[dash](50,140)(25,50,130) 
\Photon(50,200)(35,220){3}{4} 
\Text(50,135)[c]{(2)} 
\SetWidth{1.5}
\Arc[arrow,arrowpos=0.25](150,175)(25,315,225) 
\SetWidth{0.5}
\Arc(150,175)(25,225,315) 
\Arc[dash](150,140)(25,50,130) 
\Photon(150,165)(135,185){3}{4} 
\Text(150,135)[c]{(3)} 
\Text(-30,172)[l]{{\small \begin{tabular}{l}
        $V_\rho=1$\\
        \\
        ${\cal O}(p^6)$\\
        \\
        NLO
        \end{tabular}
  }}
\SetWidth{1.5}
\Arc[arrow,arrowpos=0.25](50,60)(25,0,180) 
\Arc[arrow,clockwise](50,30)(25,0,180) 
\SetWidth{0.5}
\Line[arrow](25,30)(75,30)
\Line[arrow](25,60)(75,60)
\ZigZag(25,30)(25,60){1.5}{4.0}
\ZigZag(75,30)(75,60){1.5}{4.0}
\ZigZag(40,30)(40,60){1.5}{4.0}
\ZigZag(60,30)(60,60){1.5}{4.0}
\Photon(50,85)(35,105){3}{4}
\Text(50,45)[c]{{\large $\cdots$}}
\Line[arrow,arrowpos=0.5](20,-30)(75,-30) 
\Line[arrow,arrowpos=0.5](20,-60)(75,-60)
\ZigZag(20,-30)(20,-60){1.5}{4.0}
\ZigZag(75,-30)(75,-60){1.5}{4.0}
\ZigZag(35,-30)(35,-60){1.5}{4.0}
\ZigZag(60,-30)(60,-60){1.5}{4.0}
\SetWidth{1.5}
\Line[arrow,arrowpos=1](75,-30)(20,-60)
\Line[arrow,arrowpos=1](75,-60)(20,-30)
\SetWidth{0.5}
\Photon(32,-54)(13,-35){3}{4}
\Text(47.5,-37.5)[c]{{\large $\cdots$}}
\Text(47,-76)[c]{(4)} 
\SetWidth{1.5}  
\Arc[arrow](150,60)(25,0,180) 
\Arc[arrow,clockwise](150,30)(25,0,180) 
\SetWidth{0.5}
\Line[arrow](125,30)(175,30)
\Line[arrow,arrowpos=0.25](125,60)(175,60)
\ZigZag(125,30)(125,60){1.5}{4.0}
\ZigZag(143,30)(143,60){1.5}{4.0}
\ZigZag(158,30)(158,60){1.5}{4.0}
\ZigZag(175,30)(175,60){1.5}{4.0}
\Photon(153,60)(135,75){3}{4}
\Text(135,45)[c]{{\small $\cdots$}}
\Text(168,45)[c]{{\small $\cdots$}}
\Line[arrow,arrowpos=0.5](120,-30)(175,-30) 
\Line[arrow,arrowpos=0.5](120,-60)(175,-60)
\ZigZag(125,-30)(125,-60){1.5}{4.0}
\ZigZag(143,-30)(143,-60){1.5}{4.0}
\ZigZag(158,-30)(158,-60){1.5}{4.0}
\ZigZag(175,-30)(175,-60){1.5}{4.0}
\Text(135,-45)[c]{{\small $\cdots$}}
\Text(168,-45)[c]{{\small $\cdots$}}
\Photon(153,-30)(135,-15){3}{4}
\SetWidth{1.5}
\Line[arrow,arrowpos=1](175,-30)(120,-60)
\Line[arrow,arrowpos=1](175,-60)(120,-30)
\Text(147,-76)[c]{(5)} 
\SetWidth{1.5}  
\Arc[arrow](250,60)(25,0,180) 
\Arc[arrow,clockwise](250,30)(25,0,180) 
\SetWidth{0.5}
\Line[arrow](225,30)(275,30)
\Line[arrow,arrowpos=0.65](225,60)(275,60)
\ZigZag(225,30)(225,60){1.5}{4.0}
\ZigZag(238,30)(238,60){1.5}{4.0}
\ZigZag(263,30)(263,60){1.5}{4.0}
\ZigZag(275,30)(275,60){1.5}{4.0}
\Line[dash](250,30)(250,60)
\Photon(250,50)(238,70){3}{4}
\Text(231,45)[c]{{\tiny $\cdots$}}
\Text(269,45)[c]{{\tiny $\cdots$}}
\Line[arrow,arrowpos=0.5](220,-30)(275,-30) 
\Line[arrow,arrowpos=0.5](220,-60)(275,-60)
\ZigZag(225,-30)(225,-60){1.5}{4.0}
\ZigZag(238,-30)(238,-60){1.5}{4.0}
\ZigZag(263,-30)(263,-60){1.5}{4.0}
\ZigZag(275,-30)(275,-60){1.5}{4.0}
\Line[dash](250,-30)(250,-60)
\Photon(250,-40)(238,-20){3}{4}
\Text(231,-45)[c]{{\tiny $\cdots$}}
\Text(269,-45)[c]{{\tiny $\cdots$}}
\SetWidth{1.5}
\Line[arrow,arrowpos=1](275,-30)(220,-60)
\Line[arrow,arrowpos=1](275,-60)(220,-30)
\Text(247,-76)[c]{(6)} 
\Text(-30,0)[l]{{\small \begin{tabular}{l}
        $V_\rho=2$\\
        \\
        ${\cal O}(p^6)$\\
        \\
        NLO
        \end{tabular}
  }}
\end{axopicture}
\caption{{\small
The set of diagrams for the evaluation of the quark condensate  up to NLO in 
in-medium ChPT. The scalar source is indicated by the wavy line, while the rest of lines have the 
standard meaning.}
\label{fig.180904.1}}
\end{center}
\end{figure}

To be specific, let us consider the set of diagrams in Fig.~\ref{fig.180904.1} for the evaluation of the quark condensate 
up to ${\cal O}(p^6)$ or NLO in the chiral counting of Eq.~\eqref{fff}. The former is the expectation value 
of $m_q\langle \Omega|\bar{q}_iq_j|\Omega\rangle$, where $m_q$ is the mass of a certain quark flavor, and
 $|\Omega\rangle$ represents the interacting nuclear-matter 
state (which evolves from the pure Fermi seas of protons and neutrons at asymptotic times $t\to \pm \infty$ 
 \cite{prcoller}). The diagrams (1)--(3) were already calculated in Ref.~\cite{annp} and the ones 
at the bottom of the figure were discussed by Ref.~\cite{lacour.180827.4}. 
The diagrams (4) and (5) arise from the quark mass dependence of the nucleon mass, while diagrams (6) do so 
from the quark mass dependence of the pion mass. 
It turns out that the set of diagrams (4) and (5) cancel each other. This is not only a characteristic 
of the quark condensate but also of the coupling $f_t$ to the temporal component of the axial-vector current 
\cite{lacour.180827.4} and of the pion self energy \cite{lacour.180827.2,lacour.180827.4}. 
In this way, the only remaining 
contribution involving the $NN$ interactions to the quark condensate are those from the diagrams (6), 
that involve only the long-range part of 
 the $NN$ interactions from the OPE contribution in Fig.~\ref{fig:wig}. 
 It was also discussed in Ref.~\cite{lacour.180827.4} that the Feynman-Hellmann theorem relating the value 
of the in-medium quark condensate and the derivative of the energy density of the nuclear medium with respect to 
the corresponding quark masses is fulfilled order by order.

A non-vanishing quark condensate is a sufficient condition for having chiral symmetry breaking, but it is not 
necessary. A sufficient and necessary condition for chiral symmetry breaking is that $f_t\neq 0$,
 because then $|\Omega\rangle$ is not invariant under chiral transformations \cite{annp}. 
As stated above, due to the mutual cancellation of the diagrams of types (4) and (5) in Fig.~\ref{fig.180904.1}, 
the calculations of $f_t$ in Ref.~\cite{annp,lacour.180827.4} and  the in-medium pion self energy 
in Refs.~\cite{annp,lacour.180827.3} hold up to NLO with the in-medium chiral counting of 
Eq.~\eqref{fff}. It was found in Ref.~\cite{annp} that $f_t$ decreases linearly with the density of the medium (because 
$V_\rho=1$ for the pertinent non-vanishing contributions), 
so that for symmetric nuclear matter the result found is 
\begin{align}
\label{180905.1}
f_t(\rho)&=f_\pi\left[1+\frac{2\rho}{f_\pi^2}\left(c_2+c_3-\frac{g_A^2}{8m}\right)\right]\\
&= f_\pi\left[1-\frac{\rho}{\rho_0}(0.26\pm 0.04)\right]~,\nn
\end{align}
with $\rho_0\simeq 0.16$~fm$^{-3}$, the nuclear matter saturation density. 
 This equation indicates a trend towards lower values of $f_t(\rho)$ as $\rho$ increases. 
Since $f_t(\rho)$ is an important 
 quantity to unveil the chiral properties of the nuclear medium, 
 a calculation at N$^2$LO should be pursued. 
The linear relation in Eq.~\eqref{180905.1} would imply that $f_t$ becomes zero 
for $\rho\simeq  4 \rho_0$, and therefore for a Fermi momentum of around 440~MeV, 
which is a priori too high for this equation being accurate. 

Regarding the calculation of the pion self energy, $\Pi(p)$,
this is an important problem as it is in tight 
correspondence to that of pionic atoms due to the relation between the pion self-energy 
and the pion-nucleus optical potential. 
The equivalent potential is given by 
$\Pi(p)/2\omega(\vp)$ with $\omega(\vp)=\sqrt{m_\pi^2+\vp^2}$. 
For the first time in the literature, Ref.~\cite{annp} explained the origin 
of the so-called $S$-wave missing repulsion in pionic atoms as an effect originating 
by the sensitive energy dependence of the pion self-energy around threshold, so that 
one should not use $\Pi(p^0=m_\pi,\vp)$ but keeping the variability of the energy argument, namely, 
$\Pi(p)$. 

This was shown by calculating the pion masses in the medium, which is defined in terms of the pion  
self-energy for $p^0=\widetilde{M}_\pi$ and $\vp=0$, as the solution of the pion dispersion equation, 
\begin{align}
\label{180906.1}
\widetilde{M}_\pi^2-m_\pi^2+\Pi(\widetilde{M}_\pi,\vec{0})=0~.
\end{align} 
Of course, for the case of asymmetric proton and neutron densities 
the pion masses are different depending on the pion charge.
 For the nucleus $^{207}$Pb  deeply bound $\pi^-$ states have been detected \cite{g,i} with a shift in the
effective in-medium $\pi^-$ mass of $\Delta M_{\pi^-}=23-27$ MeV \cite{i} for the 
pion-nucleus optical potentials used (with $\Delta M_{\pi^-}=\widetilde{M}_{\pi^-}-M_{\pi^-})$. 
By solving the equation for the $\pi^-$ mass in such circumstances Ref.~\cite{annp} obtains 
$\Delta M_{\pi^-}=18 \pm 5$~ MeV, which is already compatible with the previous values from 
Ref.~\cite{i}.

This important conclusion indicates that in-medium ChPT accounts for most of the required
shift in the $\pi^-$ mass at finite density from  experiments
on deeply bound pionic atoms \cite{i}. The main contribution to this shift,
around 16 MeV, results from the combination 
$1+4\hatr(c_2+c_3-g_A^2/8m_N)/f^2$ that appears multiplying $\widetilde{M}_{\pi^-}^2$ 
in the $\pi^-$ dispersion relation
\begin{align}
\label{180905.2}
&\widetilde{M}_{\pi^-}^2\left(1+\frac{4\hat{\rho}}{f_\pi^2}\big[c_2+c_3-\frac{g_A^2}{8m}\big]\right)
-\frac{\bar{\rho}\widetilde{M}_{\pi^-}}{f_\pi^2}
-M^2_{\pi^-}\left(1+c_1\frac{8\hat{\rho}}{f_\pi^2}\right)=0~,
\end{align}
where $\hat{\rho}=\frac{1}{2}(\rho_p+\rho_n)$ and 
$\bar{\rho}=\frac{1}{2}(\rho_p-\rho_n)$~.
 This factor corresponds to the
wave-function renormalization of the in-medium pions in symmetric nuclear
matter at threshold \cite{annp}, which for $\rho=\rho_0$ is $\sim 0.5$. 
Additionally, the factor multiplying $M_{\pi^-}^2$ in Eq.~\eqref{180905.2} 
is around 0.5, such that,
when divided by the wave-function renormalization constant, it is slightly
greater than one and it further increases $\widetilde{M}_{\pi^-}$ by the extra amount of 2
MeV. The addition of both contributions account for the final 18 MeV in $\Delta M_{\pi^-}$, as 
referred above \cite{annp}.

\section{Conclusions}
\label{sec:conc}

We have reviewed on the results obtained in Refs.~\cite{lacour.180827.2,lacour.180827.3,lacour.180827.4} 
 based on the approach developed in Ref.~\cite{prcoller} and on a chiral power counting in the nuclear medium 
that combines pion-mediated \cite{annp} and short-range \cite{lacour.180827.3} internucleon interactions. 
The power counting is bounded from 
below and at a given order it requires to calculate a finite number of
mechanisms, which typically implies the resummation of  an infinite string
of  two-nucleon reducible diagrams with the leading multi-nucleon ChPT
amplitudes. These resummations arise because this power counting takes into account from the 
onset the presence of enhanced nucleon propagators and it can also be applied
to multi-nucleon forces. Non-perturbative techniques that perform these resummations both in
scattering as well as in production processes are developed based on Unitary Chiral Perturbation Theory (UChPT), 
which is adapted now to the nuclear medium by implementing the new power counting. 

Using this theory we have first discussed the 
non-perturbative leading order (LO) vacuum and in-medium  
nucleon-nucleon ($NN$) interactions. These were derived in Ref.~\cite{lacour.180827.3} and applied 
to the calculation of the energy density for nuclear matter up to next-to-leading order (NLO). The energy per baryon obtained for 
neutron and symmetric nuclear matter is 
in  good agreement with other sophisticated many-body calculation employing realistic $NN$ potentials and including 
an adjustable parameter for mimicking three-nucleon interactions \cite{urbana}. 

The equation of state for neutron matter, $P(\rho)$, is derived from these results and applied to the study 
of the neutron-star properties in Ref.~\cite{llanes.180827.1}. 
The mass-radius curve is in agreement with the recent observation of a neutron star with a mass near  
$2\,M_\odot$ and with other constraints  from the GW170817 event observed by the 
LIGO and Virgo Collaborations \cite{ligo.180827.1}. 
We have reported here the agreement with the maximum mass of a neutron star and the
 radius  for a neutron star with a mass of $1.4~M_\odot$. 
 Reference~\cite{llanes.180827.1} also makes the important point of determining an upper bond 
 for the  gravitational constant $G$ in  a neutron star of 2~$M_\odot$,
 so that it cannot be more than 12\% larger compared to its value on the surface of the Earth 
(despite the gravitational field intensity is around twelve orders of magnitudes larger).

We have also reviewed about the results of Refs.~\cite{lacour.180827.4,annp} on the pion coupling $f_t$ to the 
temporal component of the axial-vector current and the in-medium quark condensate, and of 
Refs.~\cite{lacour.180827.2,lacour.180827.3,annp} on the pion self-energy. 
The quark condensate and $f_t$ are important quantities
 to characterize the chiral symmetry breaking in nuclear matter. 
It is found that the non-linear corrections  in density for $f_t$ and the pion self-energy, that would arise from the 
in-medium non-perturbative $NN$ interactions,  
vanish identically at NLO,  while for the quark condensate they are small because strong cancellations occur. 
Finally, we have discussed the important conclusion of Ref.~\cite{annp} that in-medium ChPT explains the missing $S$-wave 
repulsion of a $\pi^-$ in the nuclear medium, responsible for the rather large shift of the $\pi^-$ mass in Pb 
\cite{i}.

Contrary to other approaches extensively used in modern literature, 
the in-medium ChPT is based on a chiral power counting that takes into account the enhanced 
nucleon propagators in the medium. Furthermore, within this theory one does not invoke alien  many-body techniques for 
complementing the chiral expansion, because  the application of the chiral power counting leads 
to the corresponding non-perturbative resummations that should be performed.

\section*{Acknowledgements}
This work is partially funded by the MINECO (Spain) and EU grant  FPA2016-77313-P.
\section*{Appendices}

\appendix{}

\section{The one-pion exchange nucleon-nucleon partial waves}
\label{app:1pi}
\def\theequation{\Alph{section}.\arabic{equation}}
\setcounter{equation}{0}   

The one-pion exchange nucleon-nucleon partial waves $\cT_{JI}^{1\pi}(\bar{\ell},\ell,S)$ up to the $F-$wave 
that result from Eqs.~\eqref{1pi.sin}--\eqref{1pi.pw} are (including the amplitudes that couple different $\ell$ and $\bar{\ell}$):
\begin{align}
  \cT^{1\pi}_{01}(S,S,0) &~=~ - \frac{g_A^2}{4f_\pi^2} \frac{1}{4p^2} \biggl[ 4p^2 + m_\pi^2 \ln \frac{m_\pi^2}{m_\pi^2+4p^2} \biggr] ~,  \displaybreak[0] \\
  \cT^{1\pi}_{10}(S,S,1) &~=~ - \frac{g_A^2}{4f_\pi^2} \frac{1}{4p^2} \biggl[ 4p^2 + m_\pi^2 \ln \frac{m_\pi^2}{m_\pi^2+4p^2} \biggr] ~, \notag \displaybreak[0] \\
  \cT^{1\pi}_{10}(P,P,0) &~=~ \frac{g_A^2}{4f_\pi^2} \frac{3m_\pi^2}{8p^4} \biggl[ 4p^2 + (m_\pi^2+2p^2)  \ln \frac{m_\pi^2}{m_\pi^2+4p^2} \biggr] ~, \notag \\
  \cT^{1\pi}_{01}(P,P,1) &~=~ \frac{g_A^2}{4f_\pi^2} \frac{1}{4p^2} \biggl[ 4p^2 + m_\pi^2 \ln \frac{m_\pi^2}{m_\pi^2+4p^2} \biggr] ~, \notag \displaybreak[0] \\
  \cT^{1\pi}_{11}(P,P,1) &~=~ \frac{g_A^2}{4f_\pi^2} \frac{1}{16p^4} \biggl[ 4p^2(m_\pi^2 - 2p^2) + m_\pi^4 \ln \frac{m_\pi^2}{m_\pi^2+4p^2} \biggr] ~, \notag \displaybreak[0] \\
  \cT^{1\pi}_{21}(P,P,1) &~=~ \frac{g_A^2}{4f_\pi^2} \frac{1}{80p^4} \biggl[ 4p^2(3m_\pi^2 + 2p^2) + m_\pi^2(3m_\pi^2 + 8p^2) \ln \frac{m_\pi^2}{m_\pi^2+4p^2} \biggr] ~, \notag \displaybreak[0] \\
  \cT^{1\pi}_{21}(D,D,0) &~=~ - \frac{g_A^2}{4 f_\pi^2} \frac{m_\pi^2}{32p^6} \biggl[ 12p^2(m_\pi^2 + 2p^2) + (3m_\pi^4 + 12m_\pi^2p^2 + 8p^4) \ln \frac{m_\pi^2}{m_\pi^2+4p^2} \biggr] ~, \notag \displaybreak[0] \\
  \cT^{1\pi}_{10}(D,D,1) &~=~ - \frac{g_A^2}{4f_\pi^2} \frac{1}{16p^4} \biggl[ 4p^2(3m_\pi^2 + 2p^2) + m_\pi^2(3m_\pi^2 + 8p^2) \ln \frac{m_\pi^2}{m_\pi^2+4p^2} \biggr] ~, \notag \displaybreak[0] \\
  \cT^{1\pi}_{20}(D,D,1) &~=~ - \frac{g_A^2}{4f_\pi^2} \frac{1}{16p^6} \biggl[ 4p^2(3m_\pi^4 + 3m_\pi^2p^2 - 2p^4) + 3(m_\pi^6 + 3m_\pi^4p^2) \ln \frac{m_\pi^2}{m_\pi^2+4p^2} \biggr] ~, \notag \displaybreak[0] \\
  \cT^{1\pi}_{30}(D,D,1) &~=~ - \frac{g_A^2}{4f_\pi^2} \frac{1}{224p^6} \biggl[ 4p^2(15m_\pi^4 + 42m_\pi^2p^2 + 8p^4) + 3m_\pi^2(5m_\pi^4 + 24m_\pi^2p^2 + 24p^4) \ln \frac{m_\pi^2}{m_\pi^2+4p^2} \biggr] ~, \notag  \displaybreak[0] \\
  \cT^{1\pi}_{30}(F,F,0) &~=~ \frac{g_A^2}{4f_\pi^2} \frac{m_\pi^2}{64p^8} \biggl[ 4p^2 (15m_\pi^4 + 60m_\pi^2p^2 + 44p^4) + 3(5m_\pi^6 + 30m_\pi^4p^2 + 48m_\pi^2p^4 + 16p^4) \ln \frac{m_\pi^2}{m_\pi^2+4p^2} \biggr] ~, \notag \displaybreak[0] \\
  \cT^{1\pi}_{21}(F,F,1) &~=~ \frac{g_A^2}{4f_\pi^2} \frac{1}{480p^6} \biggl[ 4p^2(15m_\pi^4 + 42m_\pi^2p^2 + 8p^4) + 3m_\pi^2(5m_\pi^4 + 24m_\pi^2p^2 + 24p^4) \ln \frac{m_\pi^2}{m_\pi^2+4p^2} \biggr] ~, \notag \displaybreak[0] \\
  \cT^{1\pi}_{31}(F,F,1) &~=~ \frac{g_A^2}{4f_\pi^2} \frac{1}{768p^8} \biggl[ 4p^2(45m_\pi^6 + 150m_\pi^4p^2 + 48m_\pi^2p^4 - 16p^6) \notag \\
  &~~~ \quad + 3m_\pi^4(15m_\pi^4 + 80m_\pi^2p^2 + 96p^4) \ln \frac{m_\pi^2}{m_\pi^2+4p^2} \biggr] ~, \notag \displaybreak[0] \\
  \cT^{1\pi}_{41}(F,F,1) &~=~ \frac{g_A^2}{4f_\pi^2} \frac{1}{6912p^8} \biggl[ 4p^2(105m_\pi^6 + 510m_\pi^4p^2 + 560m_\pi^2p^4 + 48p^6) \notag \\
  &~~~ \quad + 3m_\pi^2(35m_\pi^6 + 240m_\pi^4p^2 + 480m_\pi^2p^4 + 256p^6) \ln \frac{m_\pi^2}{m_\pi^2+4p^2} \biggr] ~, \notag \displaybreak[0] \\
  \cT^{1\pi}_{30}(G,G,1) &~=~ - \frac{g_A^2}{4f_\pi^2} \frac{1}{1792p^8} \biggl[ 4p^2(105m_\pi^6 + 510m_\pi^4p^2 + 560m_\pi^2p^4 + 48p^6) \notag \\
  &~~~ \quad + 3m_\pi^2(35m_\pi^6 + 240m_\pi^4p^2 + 480m_\pi^2p^4 + 256p^6) \ln \frac{m_\pi^2}{m_\pi^2+4p^2} \biggr] \notag  \displaybreak[0] \\
  \cT^{1\pi}_{10}(S,D,1) &~=~ -\frac{g_A^2}{4f_\pi^2} \frac{\sqrt{2}}{16p^4} \biggl[ 4p^2(3m_\pi^2 - 2p^2) + m_\pi^2(3m_\pi^2 + 4p^2) \ln \frac{m_\pi^2}{m_\pi^2+4p^2} \biggr] ~, \notag \\
                         &~=~ \cT^{1\pi}_{10}(D,S,1) \notag \displaybreak[0] \\
  \cT^{1\pi}_{21}(P,F,1) &~=~ \frac{g_A^2}{4f_\pi^2} \frac{\sqrt{6}}{480p^6} \biggl[ 4p^2(15m_\pi^4 + 24m_\pi^2p^2 - 4p^2) + 3m_\pi^2(5m_\pi^4 + 18m_\pi^2p^2 + 8p^4) \ln \frac{m_\pi^2}{m_\pi^2+4p^2} \biggr] \notag \\
                         &~=~ \cT^{1\pi}_{21}(F,P,1) ~, \notag \displaybreak[0] \\
  \cT^{1\pi}_{30}(D,G,1) &~=~ -\frac{g_A^2}{4f_\pi^2} \frac{\sqrt{3}}{896p^8} \biggl[ 4p^2(105m_\pi^6 + 390m_\pi^4p^2 + 224m_\pi^2p^4 - 16p^6) \notag \\
  &~~~ \quad + 3m_\pi^2(35m_\pi^6 + 200m_\pi^4p^2 + 288m_\pi^2p^4 + 64p^6) \ln \frac{m_\pi^2}{m_\pi^2+4p^2} \biggr] \notag \\
                         &~=~ \cT^{1\pi}_{30}(G,D,1) ~,\notag
\end{align}

\section{Calculation of the $L_{10}^{I_3}$ function}
\label{app:l10}
\setcounter{equation}{0}

\begin{figure}
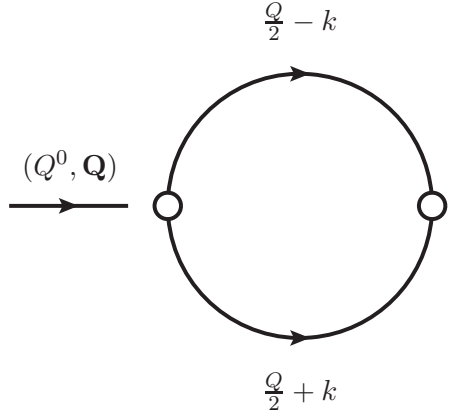

\begin{center}
\begin{axopicture}(200,175)(0,15)
\SetWidth{1.5}
\Arc[arrow,clockwise](100,100)(50,175,5) 
\Arc[arrow](100,100)(50,185,355) 
\ECirc(50,100){5} 
\ECirc(150,100){5} 
\Line[arrow](-10,100)(35,100) 
\Text(-5,115)[l]{$(Q^0,\vQ)$} 
\Text(100,170)[c]{$\frac{Q}{2}-k$} 
\Text(100,30)[c]{$\frac{Q}{2}+k$}  
\end{axopicture}
\caption[pilf]{\protect \small The 
$L_{10}$ function.
\label{fig:l10}}
\end{center}
\end{figure}


Employing the form of the nucleon propagator in Eq.~\eqref{nuc.prob}, we have for the function 
$L_{10}^{I_3}$, 
\begin{align}
L_{10}^{I_3}&=i\int\frac{d^4k}{(2\pi)^4}\left[
\frac{ \theta(\xi_1-|\val-\vk|)}
{a^0-k^0-E(\val-\vk)-i\ep}+\frac{\theta(|\val-\vk|-\xi_1)}{
a^0-k^0-E(\val-\vk)+i\ep}\right]\nn\\
&\times\left[\frac{\theta(\xi_2-|\val+\vk|)}{a^0+k^0-E(\val
+\vk)-i\ep}+\frac{\theta(|\val+\vk|-\xi_2)}{a^0+k^0-E(\val
+\vk)+i\ep}
\right]~,
\label{l10.b.f}
\end{align}
where  
\be
\vec{\alpha}=\frac{1}{2}(\vp_1+\vp_2)=\frac{\vQ}{2}~.
\label{def.alp}
\ee

Only the contributions in
Eq.~\eqref{l10.b.f} with the two poles in $k^0$ lying on opposite halves of the
complex $k^0$ plane contribute. As a result,
\begin{align}
L_{10}^{I_3}&=m\int\frac{d^3k}{(2\pi)^3}\left[
\frac{\theta(|\val-\vk|-\xi_1)\theta(|\val+\vk|-\xi_2)}{
A-\vk^2+i\ep}-\frac{\theta(\xi_1-|\val-\vk|)\theta(|\xi_2-|\val
+\vk|)}{A-\vk^2-i\ep}\right]~.
\label{l10.b.pp}
\end{align}
The first term is the particle-particle part and the last is the hole-hole
 one. Notice the different position of the cuts in $A$.
While for the particle-particle case $A$ has  a negative
imaginary part, $-i\ep$, for the hole-hole part the cut takes values with
 positive imaginary part, $+i\ep$.  The latter cut 
 requires $\vk^2=A$, but $|\vk|$ is bounded so that the two
$\theta$-functions are satisfied simultaneously. 
The requirements on $A$ are explicitly worked out in Eq.~\eqref{l10d.exp}.

The actual calculation here of the function $L_{10}$ (for simplicity we drop the superscript $I_3$) 
is given by employing the equivalent form
\begin{align}
L_{10}&=i\int\frac{d^4 k}{(2\pi)^4}\left[
\frac{1}{Q^0/2-k^0-w(\frac{\vQ}{2}-\vk)+i\epsilon}
+2\pi i \theta(\xi_1-|\frac{\vQ}{2}-\vk|) \delta(Q^0/2-k^0-w(\frac{\vQ}{2}-\vk)) \right]
\nn\\
&\times
\left[
\frac{1}{Q^0/2+k^0-w(\frac{\vQ}{2}+\vk)+i\epsilon}
+2\pi i \theta(\xi_2-|\frac{\vQ}{2}+\vk|)\delta(Q^0/2+k^0-w(\frac{\vQ}{2}+\vk))
\right]~,
\label{l10.def}
\end{align}
where we have used the expression for the nucleon propagator in Eq.~\eqref{nuc.pro}.  
This integration corresponds to the loop in Fig.~\ref{fig:l10},
where the four-momentum attached to each internal line is shown.
 The different contributions to $L_{10}$ are thus calculated according to the 
number of in-medium insertions in the nucleon propagators.

\subsection{Free part, $L_{10,f}$}
\label{app:l10f}

We perform first the $k^0$ integration by applying the Cauchy's theorem,
\begin{align}
L_{10,f}&=i\int\frac{d^4 k}{(2\pi)^4}\frac{1}{Q^0/2-k^0-w(\val-\vk)+i\epsilon}
\frac{1}{Q^0/2+k^0-w(\val+\vk)+i\epsilon}\nn\\
&=\int\frac{d^3k}{(2\pi)^3}\frac{1}{Q^0-\frac{\vk^2}{m}-\frac{\val^2}{m}+i\epsilon}=-m\int \frac{d^3 k}{(2\pi)^3}\frac{1}{\vk^2-A}\nn\\
&=
-\frac{m}{2\pi^2}\int_0^\Lambda dk-\frac{mA}{4\pi^2}\int_{-\infty}^\infty \frac{dk}{k^2-A-i\epsilon}=-\frac{m\Lambda}{2\pi^2}-\frac{im \sqrt{A}}{4\pi}~.
\label{8.22}
\end{align}
 with 
\be
A=mQ^0-\frac{\vQ^2}{4}+i\epsilon=mQ^0-\alpha^2+i\epsilon~,
\label{def.a}
\ee
and $\alpha=|\vec{\alpha}|$. One has to keep in mind in the following the $+i\epsilon$ prescription in the definition of $A$. 
In order to emphasize this, we write explicitly the combination $A+i\epsilon$  in many integrals, 
though the $+i\epsilon$ is already contained in $A$ according to Eq.~\eqref{def.a}.

The result in Eq.~\eqref{8.22}  corresponds to Eq.~\eqref{gdc}, 
as it should because $g(A)=L_{10,f}(A)$. Note that here we have used a somewhat 
different scheme of calculation starting from four dimensions and removing the 
temporal component by explicit integration, so that we end with Eq.~\eqref{int.g} afterwards.

\noindent

\subsection{One-medium insertion, $L_{10,m}$}
For the one-medium insertion, $L_{10,m}$ the $k^0$-integration is done by making use of the energy-conserving Dirac delta-function in the in-medium part of the nucleon propagator. We are then left with
\begin{align}
L_{10,m}&=-\int \frac{d^3k}{(2\pi)^3}
\frac{\theta(\xi_1-|\vk-\val|)+\theta(\xi_2-|\vk+\val|)}{Q^0-\frac{\vk^2}{m}-\frac{\val^2}{m}+i\epsilon}\nn\\
&=m\int \frac{d^3 k}{(2\pi)^3}\frac{\theta(\xi_1-|\vk-\val|)+\theta(\xi_2-|\vk+\val|)}{\vk^2-A-i\epsilon}~.
\label{l10.m.def}
\end{align}
Let us concentrate on evaluating the integral,
\begin{align}
\label{h1}
&\ell_{10,m}(\xi_1,A,\alpha)=m \int\frac{d^3k}{(2\pi)^3}\frac{\theta(\xi_1-|\vk-\val|)}{\vk^2-A-i\epsilon}\\
&=\frac{m}{4\pi^2}\left\{
\xi_1-\sqrt{A} \hbox{ arctanh}\frac{\xi_1-\alpha}{\sqrt{A}}
-\sqrt{A}\hbox{ arctanh}\frac{\xi_1+\alpha}{\sqrt{A}}    - \frac{A+\alpha^2-\xi_1^2}{4\alpha}
\log\frac{(\alpha+\xi_1)^2-A}{(\alpha-\xi_1)^2-A}
\right\}~.\nn
\end{align}
Here we have taken into account that the Heaviside function in the numerator implies the conditions,
\begin{align}
&\alpha\geq \xi_1~,\nn \\
&|\vk|\in[\alpha-\xi_1,\alpha+\xi_1]~,~\cos\theta\in 
\left[\frac{\vk^2+\alpha^2-\xi_1^2}{2|\vk|\alpha},1\right]~.\nn\\
&\alpha<\xi_1~,\nn\\
&|\vk|\in [0,\xi_1-\alpha]~,~\cos\theta\in[-1,1]~,\nn\\
&|\vk|\in [\xi_1-\alpha,\xi_1+\alpha]~,~\cos\theta\in \left[\frac{\vk^2+\alpha^2-\xi_1^2}{2|\vk|\alpha},1
\right]~.
\label{1m.con}
\end{align}
Despite the  separation between the cases $\alpha\geq \xi_1$ and $\alpha<\xi_1$, both give rise to the
same expression in Eq.~\eqref{h1}. In terms of the function $\ell_{10,m}(\xi_1,A,\alpha)$, Eq.~\eqref{l10.m.def}, one has 
\be
L_{10,m}(\xi_1,\xi_2,A,\alpha)=\ell_{10,m}(\xi_1,A,\alpha)+\ell_{10,m}(\xi_2,A,\alpha)~.
\ee

\subsection{Two medium insertions, $L_{10,d}$}
\label{app:l10.d}
For the case with two medium insertions
\begin{align}
L_{10,d}&=\frac{-i}{(2\pi)^2}\int d^4 k\, \theta(\xi_1-|\vk-\val|)\theta(\xi_2-|\vk+\val|) 
\delta(\frac{Q^0}{2}-k^0-w(|\vk-\val|))\delta( \frac{Q^0}{2}+k^0-w(\vk+\val))\nn\\
&=\frac{-im \sqrt{A}}{ 8\pi^2}\int d\hat{\vk} \, \theta(\xi_1-|\hat{\vk} \sqrt{A}-\val|)\theta(\xi_2- |\hat{\vk}\sqrt{A}
+\val|)
\label{l10d.def}
\end{align}
We take in the following that $\xi_2\geq \xi_1$. 
If the opposite were true one can use the same expressions that we derive below but with the exchange 
$\xi_1\leftrightarrow \xi_2$. This is clear after changing $\hat{\vk}\to -\hat{\vk}$  in the integral of Eq.~\eqref{l10d.def}. 

The two step functions can be easily solved. Denoting by $\theta$ the angle between $\hat{\vk}$ and 
$\val$, they imply
\begin{align}
\cos\theta &\geq \frac{A+\alpha^2-\xi_1^2}{2\alpha\sqrt{A}}\equiv y_1 \nn\\
\cos\theta &\leq \frac{\xi_2^2-A-\alpha^2}{2\alpha\sqrt{A}} \equiv y_2~.
\label{l10d.con}
\end{align}
One has to require that $y_1\leq 1$ and that $y_2\geq -1$, otherwise $\cos\theta$ is out of the range 
$[-1,+1]$ from the 
conditions $(\ref{l10d.con})$. In addition, it is also necessary that $y_2\geq y_1$. 
\begin{align}
y_1\leq +1  & \to   \alpha-\xi_1 \leq \sqrt{A} \leq \alpha+\xi_1~,\nn\\
y_2\geq -1 & \to  \alpha-\xi_2\leq \sqrt{A} \leq \alpha+ \xi_2~,\nn\\
y_1\leq y_2& \to  A\leq \frac{\xi_1^2+\xi_2^2}{2}-\alpha^2\equiv A_{max}~.
\label{l10d.con1}
\end{align}
For $\alpha\geq \xi_1$ in order that $(\alpha-\xi_1)^2\leq A_{max}$, as the last of the three previous conditions requires, then 
\be
\alpha\leq \frac{\xi_1+\xi_2}{2}~.
\label{l10d.alp.con}
\ee
Notice that because $\xi_2\geq \xi_1$ the previous upper bound is larger than $\xi_1$. 
From Eq.~\eqref{l10d.alp.con} it follows then that $\alpha- \xi_2\leq 0$. In addition it is always the case that 
$(\alpha+\xi_2)^2\geq A_{max}$. On the other hand, 
\begin{align}
\hbox{if ~} \alpha\geq \frac{\xi_2-\xi_1}{2}\to A_{max}\leq (\alpha+\xi_1)^2~,\nn\\
\hbox{if ~} \alpha\leq \frac{\xi_2-\xi_1}{2}\to A_{max} \geq (\alpha+\xi_1)^2~.
\end{align}
For the final form of $L_{10,d}$ one also has to take into account the conditions,
\begin{align}
y_1\geq -1 &\to \sqrt{A}\geq \xi_1-\alpha,\nn\\
y_2 \leq +1 &\to  \sqrt{A}\geq \xi_2-\alpha~.
\label{l10d.con2}
\end{align}
Gathering together the conditions in Eqs.~\eqref{l10d.con}--\eqref{l10d.con2} we have the following options,
\be
 y_1\leq -1~,~y_2\leq +1 ~\to ~ \xi_2-\alpha\leq \sqrt{A}\leq \xi_1-\alpha~,
\ee
which is not possible because $\xi_2\geq \xi_1$.
\be
y_1\leq -1~,~y_2\geq +1~\to ~\sqrt{A}\leq \xi_1-\alpha~,
\label{l10d.cos.1}
\ee
this only holds for $\alpha\leq \xi_1$. Then $\cos\theta\in [-1,+1]$ and $L_{10,d}=-i\, m\sqrt{A}/(2 \pi)$.

\be
-1\leq y_1 \leq +1~,~ y_2\geq +1~\to ~ |\xi_1-\alpha|\leq \sqrt{A}\leq \hbox{min}(\xi_1+\alpha,\xi_2-\alpha)~,  
\label{l10d.cos.2}
\ee
in which case, $\cos\theta\in [y_1,+1]$ and $L_{10,d}=-im(\xi_1^2-(\sqrt{A}-\alpha)^2)/(8\pi\alpha)$~. It follows that 
 $\xi_1+\alpha\leq \xi_2-\alpha$ for $\alpha\leq (\xi_2-\xi_1)/2$ and $\xi_1+\alpha\geq \xi_2-\alpha$ for 
$\alpha\geq (\xi_2- \xi_1)/2$. In both cases $[{\rm min}(\xi_1+\alpha,\xi_2-\alpha)]^2\leq 
 A_{max}$, as can be easily seen.
 
The last possibility is that
\be
-1\leq y_1 \leq +1~,~y_2\leq +1~\to~\xi_2-\alpha\leq \sqrt{A}\leq \xi_1+\alpha~.
\label{l10d.cos.3}
\ee
For this case to hold, it is necessary that $\alpha \geq (\xi_2-\xi_1)/2$. But then $A_{max}\leq (\xi_1+\alpha)^2$ so that
the allowed upper limit for $\sqrt{A}$ is $\sqrt{A_{max}}$ not $\xi_1+\alpha$. In this case, $\cos\theta\in [y_1,y_2]$ and  
$L_{10,d}=-i \,m(\xi_1^2+\xi_2^2-2 A-2\alpha^2)/(8\pi\alpha)$.

In summary,
\be
L_{10,d}=\left\{
\begin{array}{ll}
 -\frac{i\, m \sqrt{A}}{2\pi}~,& \sqrt{A}\leq \xi_1-\alpha~,~\alpha\leq \xi_1 \\
-\frac{i\, m}{8\pi\alpha}(\xi_1^2-(\sqrt{A}-\alpha)^2)~, & |\xi_1-\alpha|\leq \sqrt{A} \leq \xi_1+\alpha~,~\alpha\leq \frac{\xi_2-\xi_1}{2}  \\
-\frac{i\, m}{8\pi\alpha}(\xi_1^2-(\sqrt{A}-\alpha)^2)~, & |\xi_1-\alpha|\leq \sqrt{A} \leq \xi_2-\alpha~,~\frac{\xi_2-\xi_1}{2}\leq \alpha \leq \frac{\xi_1+\xi_2}{2}  \\
-\frac{i\, m}{8\pi\alpha}(\xi_1^2+\xi_2^2-2 A-2\alpha^2)~, & \xi_2-\alpha\leq \sqrt{A} \leq \sqrt{A_{max}}~,~\frac{\xi_2-\xi_1}{2}\leq \alpha \leq \frac{\xi_1+\xi_2}{2} ~.
\end{array}
\right.
\label{l10d.exp}
\ee


\end{document}